\documentclass[aps,english,prd,showpacs,showkeys,twocolumn,longbibliography,nofootinbib]{revtex4-1}

\usepackage[english]{babel}
\usepackage[utf8]{inputenc}
\usepackage[T1]{fontenc}
\usepackage{fnpct} 
\usepackage{ulem} 
\usepackage{scrextend} 

\usepackage{amsmath}
\usepackage{amsfonts}
\usepackage{amssymb}
\usepackage{empheq}
\usepackage{tensor}

\usepackage{fancyvrb}
\usepackage{graphicx}
\usepackage{floatrow}
\usepackage[usenames,dvipsnames]{xcolor}

\usepackage{url}
\usepackage[pdftex,colorlinks=true, pdfstartview=FitV, linkcolor= Mlink, citecolor=Mlink, urlcolor= Mlink, hyperindex=true, hyperfigures=true]{hyperref}
\hypersetup{linktoc=page}

\definecolor{Mlink}{rgb}{0.8, 0.25, 0.33}
\definecolor{Mred}{rgb}{0.7,0,0}
\definecolor{Mgreen}{rgb}{0, 0.7, 0}
\definecolor{Mblue}{rgb}{0, 0, 0.7}
\definecolor{Mgray}{rgb}{0.25, 0.25, 0.25}

\definecolor{LinkJournal}{rgb}{0.7,0,0}
\definecolor{LinkADS}{rgb}{0.2,0,0.7}
\definecolor{LinkArXiv}{rgb}{0.1,0.5,0.1}

\newcommand{\proof}[1]{\noindent \textit{Proof. }{ #1 }}
\newcommand{\remark}[1]{\noindent \textit{Remark. }{#1 \\ }}
\newcommand{\definition}[1]{\noindent \textit{Definition. }{#1 \\ }}

\newcommand{\dd}{\ensuremath{\mathrm{d}}}

\newcommand{\T}[1]{\boldsymbol{#1}}
\newcommand{\Tt}[1]{\bar{\boldsymbol{#1}}}
\newcommand{\Tb}[1]{\text{\b{$\boldsymbol{#1}$}}}

\newcommand{\Tbb}[1]{\text{\b{\b{$\boldsymbol{#1}$}}}}
\newcommand{\Lie}[1]{\CL_{#1}}

\newcommand{\CM}{\mathcal{M}}

\newcommand{\CQ}{\mathcal{Q}}

\newcommand{\CL}{\mathcal{L}}

\newcommand{\vN}{v}

\newcommand{\D}{D}
\newcommand{\V}{V}
\newcommand{\U}{U}

\newcommand{\derivt}[2]{\tensor[^{#2}]{\partial}{_{t|_{#1}}}}
\newcommand{\Tderivt}[2]{\tensor[^{#2}]{\T\partial}{_{t|_{#1}}}}
\newcommand{\derivtN}[1]{\derivt{#1}{}}
\newcommand{\derivtn}[1]{\derivt{#1}{\T n}}
\newcommand{\Tderivtn}[1]{\Tderivt{#1}{\T n}}

\newcommand{\SNew}{\Sigma^\mathrm{N}}
\newcommand{\MNew}{\mathcal{M}^\mathrm{N}}
\newcommand{\folGR}{\{\Sigma_t\}_{t\in \mathbb{R}}}
\newcommand{\folNew}{\{\SNew_t\}_{t\in \mathbb{R}}}

\newcommand{\vt}{w}
\newcommand{\B}{{\mathcal{B}}}
\newcommand{\Nfol}{\mathcal{N}}

\newcommand{\class}[2]{{\mathcal{X}_{#1}^{#2}}}
\newcommand{\classn}[1]{{\mathcal{X}_{#1}^{\T n}}}
\newcommand{\classN}[1]{{\mathcal{X}_{#1}^{}}}

\newcommand{\Riem}{\tensor[^4]{R}{}}

\newcommand{\TuN}{\tensor[^{\mathrm{N}}]{\T u}{}}
\newcommand{\uN}{\tensor[^{\mathrm{N}}]{u}{}}

\newcommand{\TDh}{\hat{\T\D}}
\newcommand{\Dh}{\hat{\D}}
\newcommand{\TRh}{\hat{\T R}}
\newcommand{\Rh}{\hat{R}}
\newcommand{\hh}{\hat{h}}
\newcommand{\Thh}{\hat{\T h}}

\newcommand{\convP}{\tilde}
\newcommand{\xP}{\convP{x}}
\newcommand{\DP}{\convP{\D}}
\newcommand{\vNP}{\convP{\vN}}
\newcommand{\VP}{\convP{\V}}
\newcommand{\UP}{\convP{\U}}
\newcommand{\gP}{\convP{g}}

\newcommand{\parP}{\convP{\partial}}
\newcommand{\hP}{\convP{h}}

\newcommand{\TP}{\convP{T}}

\newcommand{\convG}{}
\newcommand{\xgal}{\convG{x}}
\newcommand{\ygal}{\convG{y}}

\newcommand{\Ugal}{\convG{\U}}

\newcommand{\pargal}{\convG{\partial}}
\newcommand{\hgal}{\convG{h}}
\newcommand{\nabgal}{\convG{\nabla}}

\newcommand{\Tgal}{\convG{T}}

\input{aastexv6.sty} 
\usepackage{filecontents}

\begin{document}

\title{1+3 formulation of Newton's equations}
\author{Quentin Vigneron}
\email{quvigneron@gmail.com}
\affiliation{Univ Lyon, Ens de Lyon, Univ Lyon1, CNRS, Centre de Recherche Astrophysique de Lyon UMR5574, F–69007, Lyon, France}

\date{\today}

\begin{abstract}
We present in this paper a 4-dimensional formulation of the Newton equations for gravitation on a Lorentzian manifold (hence distinct from the Newton-Cartan formalism), inspired from the 1+3 and 3+1 formalisms of general relativity.
We begin by writing the Newton equations in a general time-parametrised coordinate system. We show that the freedom on the coordinate velocity of this system with respect to a Galilean reference system is similar to the shift freedom in the 3+1-formalism of general relativity. This allows us to write Newton's theory as living in a 4-dimensional Lorentzian manifold $\MNew$. This manifold can be chosen to be curved depending on the dynamics of the Newtonian fluid. In this paper, we focus on a specific choice for $\MNew$ leading to what we call the \textit{1+3-Newton equations}. We show that these equations can be recovered from general relativity with a Newtonian limit performed in the rest frames of the relativistic fluid.
The 1+3 formulation of the Newton equations along with the Newtonian limit we introduce also allow us to define a dictionary between Newton's theory and general relativity. This dictionary is defined in the rest frames of the dust fluid, i.e. a non-accelerating observer. A consequence of this is that it is only defined for irrotational fluids.
As an example supporting the 1+3-Newton equations and our dictionary, we show that the parabolic free-fall solution in 1+3-Newton exactly translates into the Schwarzschild spacetime, and this without any approximations. The dictionary might then be an additional tool to test the validity of Newtonian solutions with respect to general relativity. It however needs to be further tested for non-vacuum, non-stationary and non-isolated Newtonian solutions, as well as to be adapted for rotational fluids.
One of the main applications we consider for the 1+3 formulation of Newton's equations is to define new models suited for the study of backreaction and global topology in cosmology.
\end{abstract}


\maketitle

%

\section{Introduction}
\label{sec::Intro}

Fluid dynamics in Newton's and Einstein's theories of gravitation are known to be closely related on a formal aspect. This is well presented by Ellis~\cite{1971_Ellis} where the parallel between the fluid kinematical quantities and equations in Newton's theory with those in general relativity (defined via the 1+3-formalism of general relativity) is drawn. This parallel highlights the similarities, but also the differences between both theories. From the point of view of general relativity (hereafter GR) these differences appear as missing physical phenomena in the Newtonian theory. For instance we can mention the precession of perihelion for elliptic orbits or the gravitational waves. However, there also are phenomena in Newton's theory which are not included in GR. This is the case for shear-free dust solutions, which can both expand and rotate in Newton's theory, but cannot in GR (e.g. \cite{1967_Ellis}). Newtonian gravitation is then not a reduction of Einstein's theory of gravitation.

Another phenomenon which is not described by Newton's theory is the backreaction of inhomogeneities on the global expansion of the Universe. This effect was shown to be exactly zero for a compact space in Newton by Buchert \& Ehlers~\cite{1997_Buchert_et_al}. Though no such theorem exists in GR, we still do not know whether or not the backreaction might play a major role in the expansion of space: this is the \textit{backreaction problem in cosmology}.

Then, due to Buchert~\&~Ehlers~\cite{1997_Buchert_et_al}, one has to use GR to study this problem. This can be done with exact analytical classes of solutions such as the Lemaitre-Tolman-Bondi model or the Szekeres model. Though solutions to these models can be refined to feature complex structures \cite{2016_Sussman_et_al}, they suffer from highly symmetric conditions which implies that they are poorly representative of the Universe. That is why the last decade has seen the development of fully relativistic simulations, solving exactly the Einstein equations in a cosmological setup (e.g.~\cite{2016_Giblin_et_al_b, 2019_Macpherson_et_al}). But while these simulations were performed to probe the backreaction, for resolution reasons they still need to be realised in a cubic box with a size smaller than the Hubble radius. Like the symmetric conditions in the analytical models, the box in these simulations might act as a restrictive condition. The latter can however be physically meaningful as it corresponds to imposing a specific compact topology and size to the Universe, corresponding to the 3-torus $\mathbb{T}^3$ for these simulations.

As of today, no clear evidence for a non-trivial global topology exists. Nonetheless, studying the potential effects of different compact topologies on structure formations and backreaction remains important. This is heavy to implement in numerical simulations as most of the algorithms and frameworks require a cubic box with Cartesian coordinates\footnote{Brown~\cite{2009_Brown} might give the best formulation of the BSSN formalism to enable numerical simulations in non-flat topologies.}. One might then want to seek for an analytical GR based model which enables realistic structure formation along with the study of topological effects on backreaction. The relativistic Lagrangian perturbation theory of Ref.~\cite{RZA1} allows for realistic non-linear structure formation in the context of GR. However the backreaction in a compact space cannot be studied as the model is based on a perturbation around a homogeneous global expansion. \\

The aim of this paper is to introduce a formulation of Newton's theory which could help in the construction of such a model. This formulation is a 4-dimensional (hereafter 4D) covariant writing of Newton's equations for dust fluid dynamics on a Lorentzian manifold\footnote{We explain in Sec.~\ref{sec::sig_New} why writing Newton's equations on a Lorentzian manifold is not in contradiction with this theory.}. We call the formulation \textit{1+3-Newton}. The corresponding equations are equivalent to the classical Newton's equations. Being written on such a Lorentzian manifold the theory can then be easily compared with GR. This will allow us to better understand why some physical gravitational phenomena like the backreaction are not present in Newton. In particular we think that the 1+3-Newton formulation can be a starting point to define new simple models suited for the study of backreaction and global topology in cosmology. This will be briefly discussed at the end of this paper. We will rather focus on the construction of the new formulation of Newton's equations. \\

To support the 1+3-Newton formulation, we will show that it can be directly recovered from GR with a Newtonian limit. This is however done only for irrotational fluids. Furthermore, as a consequence of this limit, we are also able to construct a \textit{dictionary} between the Newtonian fluid kinematical quantities and the ones of GR, for vorticity-free flows. This will be an additional tool in the scope of defining new cosmological models. The observer with respect to which this dictionary is defined is the dust fluid itself, and thus is a non-accelerating observer. To the best of our knowledge this differs from existing Newton-GR dictionaries (e.g. Green~\&~Wald~\cite{2012_Green_et_al}). In this sense it might be an interesting complementary test to assess at which point non-linear Newtonian simulations are physically relevant in a cosmological context. For instance, we would complement studies like that of East et al.~\cite{2018_East_et_al} which compared Newtonian and relativistic simulations of a simplified cosmological setup using the dictionary of Green~\&~Wald~\cite{2012_Green_et_al}.

We will show that our dictionary is coherent in the specific case of the Schwarzschild spacetime. In particular we will show that the Schwarzschild manifold is the exact translation of a solution of the 1+3-Newton equations. \\

This 1+3-Newton formulation is different from the Newton-Cartan (hereafter NC) theory (e.g. K\"unzle~\cite{1976_Kunzle}) which also provides a 4D formulation of Newton's theory. The 4D-manifold in NC is however not a Lorentzian manifold as in GR but a Galilei manifold with two degenerate spacetime metrics\footnote{No non-degenerate spacetime metric is defined in NC.}. Thus this manifold does not provide a direct comparison with Einstein's theory of gravitation. In place the correspondence between both theories is given by the Newtonian limit in the frame theory of Ehlers~\cite{2019_Ehlers}. This limit however only constructs a Galilei manifold from a solution of the Einstein equations, but not the contrary. The correspondence between both theories then only works in one way, i.e. from GR to NC. In our formulation, we are able to define a non-degenerate spacetime metric, with signature $(- + + +)$, hence implying the manifold to be Lorentzian. \\

This paper will detail the construction of the 4D-Newton equations from the classical formulation of Newton's theory on a 3-dimensional (hereafter 3D) flat manifold. To do so, we firstly derive the Newton equations in a general time-parametrised coordinate system in Sec.~\ref{sec::Newton}. This first step aims at showing that the classical 3D formulation of Newton can already be covariantly written for any coordinate system. These covariant Newton equations, while still defined on a 3D-manifold, will feature similarities with the Einstein equations: in addition to the formal equivalence between the Newton equations and the 1+3-Einstein equations (well presented by Ellis~\cite{1971_Ellis}), we show that the freedom on the general coordinate system we introduced behaves like the shift freedom of the 3+1-formalism of GR.
This allows us to write the Newton equations on a 4D-manifold $\MNew$. This is presented in Sec.~\ref{sec::4D_Newt}: in this section we first review the construction of the 3+1-Einstein equations in Sec.~\ref{sec::3+1_GR}, then detail the construction of the 4D-Newton equations in Sec.~\ref{sec::4D_Newton}.

The manifold $\MNew$ needs to have what we call a \textit{Newtonian foliation} (defined in Sec.~\ref{sec::4D_Newton}). But apart from this constraint, $\MNew$ can be any 4D (pseudo)-Riemannian manifold (Riemannian or Lorentzian) and in general its properties do not depend on the dynamics of the fluid. We can however restrict $\MNew$ to depend on this dynamics. Such a restriction is studied in Sec.~\ref{sec::Ortho} and leads to what we call the \textit{1+3-Newton equations}. In this section we also present the Newtonian limit allowing us to recover these equations from GR (Sec.~\ref{sec::dic_phys}). Our Newton-GR dictionary is defined in Sec.~\ref{sec::dico} and an application is studied in Sec.~\ref{sec::Scwharz} in the case of the Schwarzschild geometry. We show that we can recover the full Schwarzschild spacetime from an exact solution of the Newton equations.

In Sec.~\ref{sec::Disc} we discuss some aspects of the 1+3-Newton equations and the related dictionary. In particular, we present some ideas, in Sec.~\ref{sec::Dic_vort}, to allow the dictionary to be physically valid for rotational flows. Section~\ref{sec::Post_New} aims at discussing the potential use of the 1+3-Newton formalism to define modified Newtonian models based on GR. We focus on the possibility to define models suited for the study of the backreaction problem and the global topology in cosmology. We, however, leave the precise definition of such models for a further study.

\section{The classical Newton system of equations}
\label{sec::Newton}

In this section, after recalling the usual form of the Newton equations, we will express them in a general time parametrised coordinate system. Then specific choices of coordinates and their interpretation will be made. Similarities with the 3+1-Einstein equations will appear. This will allow us to extend the definition of the Newton equations to 4D-spacetime manifolds in Sec.~\ref{sec::4D_Newt}.

\subsection{Notations}

In this section we define notations which will be used in the remainder of this paper. \\

Unless otherwise stated the light speed $c$ is taken to be $1$. \\

A tensor of any type, except scalars, will be denoted in bold (example: $\T g$). In the case where the type is of importance, a tensor of type $(n,m)$ will be denoted as a bold letter with $n$ over-bars and $m$ under-bars (example: $\Tbb{g}$ for a type $(0,2)$ tensor). \\

We define the symmetric part $T_{(ab)}$, the antisymmetric part $T_{[ab]}$ and the symmetric traceless part $T_{\langle ab \rangle}$ of a rank-2 tensor $\T T$ as
\begin{align*}
	&T_{(ab)} := \frac{1}{2}\left(T_{ab} + T_{ba}\right) \quad ; \quad T_{[ab]} := \frac{1}{2}\left(T_{ab} - T_{ba}\right) ; \\
	 &T_{\langle ab \rangle} := T_{(ab)} - \frac{T}{N}g_{ab},
\end{align*}
where $\T g$ is the metric of the manifold on which $\T T$ is defined and $N$ the dimension of this manifold. \\

The Lie derivative on a manifold $\CM$ of a tensor $\T T$ along a vector field $\Tt A$ is denoted $\Lie{\T A} \T T$. The Lie derivative does not commute with the metric, so for instance, for a rank-1 tensor $\T B$,  $\Lie{\T A} \Tt B \not = \Lie{\T A} \Tb B$. We will then use $\Lie{\T A} B^a$, respectively $\Lie{\T A} B_a$, to denote the coordinate components of $\Lie{\T A} \Tt B$, respectively $\Lie{\T A} \Tb B$.

Then for a vector $\T A$ and a tensor $\T T$ on a manifold $\CM$, we have
\begin{align}
	\tensor[]{{\Lie{\T A}}}{} &{T^{a_1 ...}}_{b_1...} := A^c\nabla_c \tensor{T}{^{a_1}^{...}_{b_1}_{...}} \label{eq::Lie_def} \\
		& + \sum_i {T^{a_1 ...}}_{... \underset{\underset{i}{\uparrow}}{c} ...} \nabla_{b_i} A^c - \sum_j {T^{... \overset{\overset{j}{\downarrow}}{c} ...}}_{b_1 ...} \nabla_{c} A^{a_j}, \nonumber
\end{align}
where $\T \nabla$ is the Levi-Civita connection of $\CM$.

Finally, indices running from 0 to 3 will be denoted by Greek letters ($\alpha$, $\beta$, $\gamma$, ...) and indices running from 1 to 3 by Roman letters ($a$, $b$, $c$, ...).

\subsection{General form of the Newton system}
\label{sec::Euler-Newton}

\subsubsection{In fixed coordinates}
\label{sec::Euler-Newton_fixed}

We only consider dust fluids, implying the pressure and the non-ideal fluid terms to be zero.

The Newton system of equations describes the time evolution of a fluid characterised by a scalar field $\rho$, the fluid density, and a vector field $\T \vN$, the fluid velocity. These two tensors are defined in a 3-dimensional flat manifold\footnote{The only requirement on $\SNew$ is to be flat, i.e. its Riemann tensor is zero. $\SNew$ is however not necessarily $\mathbb R^3$. It can have any topology depending on geometric compatibility conditions (see Sec.~\ref{sec::Hubble_flow_New}).} denoted $\SNew$ and are parametrised by the time $t$. They are thus function of $t$ and the position on $\SNew$. The metric on $\SNew$ is denoted $\T h$. The system of equations is composed of two evolution equations, one for the scalar $\rho$ and one for the vector $\T\vN$, and two constraint equations. Given a fixed coordinate basis vector $\{\convG{\T{ e}}_i\}_{i=1,2,3}$ on $\SNew$, i.e. the vectors $\T{e}_i$ are not parametrised by time, the evolution equations in the corresponding coordinate system $\{x^i\}_{i=1,2,3}$ are
\begin{itemize}
	\item the \textit{mass conservation equation}
		\begin{align}
			\left(\partial_{t|_{_{x}}} + \vN^kD_k\right) \rho = - \rho D_k \vN^k, \label{eq::Cont_ENew}
		\end{align}
	\item the \textit{Euler equation}
		\begin{align}
			\left(\partial_{t|_{_{x}}}  + \vN^kD_k\right)  \vN^i = g^i, \label{eq::Euler_ENew}
		\end{align}
\end{itemize}
where $D_i$ are the components of the Levi-Civita connection on $\SNew$ in the coordinates $\xgal^i$ and $\T g$ is the gravitational vector field constraint by the following equations:
\begin{itemize}
	\item the \textit{Newton-Gauss equation}
		\begin{align}
			D_k g^k = -4\pi G\rho + \Lambda, \label{eq::Ray_ENew}
		\end{align}
	\item the \textit{Newton-Faraday equation}
		\begin{align}
			D_{[i}g_{j]} = 0. \label{eq::Wg_ENew}
		\end{align}
\end{itemize}
 with $\Lambda$ the cosmological constant.
 
Due to the equivalence principle, the Euler equation~\eqref{eq::Euler_ENew} can be seen as a definition of the gravitational vector field. Then apart from this equation, the Newton system can be written independently of $\T g$. To do so, we introduce the expansion tensor $\T \Theta$ and the vorticity tensor $\T \omega$ of the vector field $\T v$ being respectively the symmetric and the antisymmetric part of the velocity gradient $\T D \T v$, with\footnote{Here we adopt the sign convention $\omega_{ij} := +D_{[i}\vN_{j]}$. This implies the relation: $\mathrm{curl} \, \vN_i = \epsilon_{ijk}\omega^{jk}$ where $\epsilon_{ijk}$ is the Levi-Civita tensor. The inverse relation is $\omega_{ij} = \frac{1}{2} \epsilon_{ijk} \mathrm{curl} \, v^k$.}
\begin{align}
	\Theta_{ij} := D_{(i} \vN_{j)} \quad ; \quad \omega_{ij} := D_{[i}\vN_{j]}, \label{eq::def_theta_omega_New}
\end{align}
and we note the trace ${\Theta_k}^k =: \theta$. The indices are lowered and raised by the metric $\T h$. We can then rewrite Eqs.~\eqref{eq::Cont_ENew}, \eqref{eq::Ray_ENew} and \eqref{eq::Wg_ENew} respectively as
\begin{align}
	&\left(\partial_{t|_{_{x}}} + \Lie{\T v}\right) \rho		= - \rho \theta, \label{eq::Cont_New} \\
	&\left(\partial_{t|_{_{x}}} + \Lie{\T v}\right) \theta		= -4\pi G \rho + \Lambda - \Theta_{ij}\Theta^{ij} + \omega_{ij}\omega^{ij}, \label{eq::Ray_New} \\
	&\left(\partial_{t|_{_{x}}} + \Lie{\T v}\right)\omega_{ij}	= 0. \label{eq::Wg_New}
\end{align}
The gravitational vector field $\T g$ is defined as
\begin{align}
	g^i := \left(\partial_{t|_{_{x}}} + \Lie{\T v}\right)v^i + v^k\left({\Theta_k}^i + {\omega_k}^i\right). \label{eq::g_def}
\end{align}
Introducing the Lie derivative in this last equation allows us to have the same differential operator acting on $\rho$, $\Theta_{ij}$ and $\omega_{ij}$.

The system~\eqref{eq::def_theta_omega_New}-\eqref{eq::Wg_New} is closed and equivalent to the system~\eqref{eq::Cont_ENew}-\eqref{eq::Wg_ENew}.

In the form~\eqref{eq::def_theta_omega_New}-\eqref{eq::Wg_New}, the Newton system is composed of 3 evolution equations for the density, the expansion and the vorticity tensors. Equations~\eqref{eq::Cont_New}, \eqref{eq::Ray_New} and \eqref{eq::Wg_New} are respectively called the \textit{Newton mass conservation equation}, the \textit{Newton-Raychaudhuri equation} and the \textit{Newton vorticity equation}. As stated above, they are valid in a fixed coordinate system (see Sec.~\ref{sec::Euler-Newton_discu} for more details). While the expansion and vorticity tensors are explicitly covariant under any change of coordinates, parametrised by time or not, the differential operator $\partial_{t|_{_{x}}}$ is not. In the next section we will see how it changes as function of the time parametrisation of the coordinate transformation. This will allow us to write the Newton system for any time parametrised coordinate system. 

\subsubsection{In general parametrised coordinates -- $\T \vN$ description}
\label{sec::Euler-Newton_Pv}

We consider a coordinate vector basis $\{\convP{\T{e}}_a\}_{a=1,2,3}$ on $\SNew$. If the vectors $\convP{\T{e}}_a$ are parametrised by time, the coordinate system they define is called a \textit{parametrised coordinate system}. We consider such a coordinate system on $\SNew$ and note it $\{\xP^a\}_{a=1,2,3}$. For this section, any component of a tensor in the fixed coordinates $x^i$ will use the Roman letters $i$, $j$, $k$, $l$, etc (example: $T_{ij}$) and the same applies for the partial derivatives with $\pargal_t := \partial_{t|_{_{\xgal}}}$ and $\pargal_i := \partial_{\xgal^i}$; any component of a tensor in the parametrised coordinates $\xP^a$ will be denoted with a tilde and will use the Roman letters $a$, $b$, $c$, $d$, etc (example: $\TP_{ab}$) and the same applies for the partial derivatives with $\parP_t := \partial_{t|_{_{\xP}}}$ and $\parP_a := \partial_{\xP^a}$.

To be able to write the Newton equations in the $\xP^a$ coordinates from the equations in the $x^i$ coordinates, we need to consider the coordinate transformation between $\xgal^i$ and $\xP^a$. This allows us to write $\xP^a$ as functions of $\xgal^i$ and $t$, and inversely $\xgal^i$ as functions of $\xP^a$ and $t$. The Jacobian matrix ${J^i}_a$ of this transformation, and its inverse ${J_i}^a$, are then
\begin{equation}
	{J^i}_a := \parP_a \xgal^i \quad ; \quad {J_i}^a := \pargal_i\xP^a. \nonumber
\end{equation}
Because the change of coordinates $\xgal^i \rightarrow \xP^a$ depends on time, in general the Jacobian will also depend explicitly on time.

The components ${\TP^{ab...}}_{\quad \ \ cd...}$ of any tensor $\T T$ in $\SNew$ are related to the components ${\Tgal^{ij...}}_{kl...}$ of that same tensor by
\begin{equation}
	{\TP^{ab...}}_{\quad \ \ cd...} := \left({J_i}^a{J_j}^b \, ... \right) {\Tgal^{ij...}}_{kl...} \left({J^k}_c{J^l}_d \, ...\right). \nonumber
\end{equation}

We consider now a tensor $\T W$ whose components in the $\{\convG{\T{ e}}_i\}_{i=1,2,3}$ basis are ${\convG W^{ij...}}_{kl...} := \partial_t {\Tgal^{ij...}}_{kl...}$ with $\T T$ a parametrised tensor. As mentioned in the previous section, because the derivative $\partial_t$ is not explicitly covariant under the change of coordinates $\xgal^i \rightarrow \xP^a(t,\xgal^i)$, the relation ${\convP W^{ab...}}_{\quad \ \ cd...} = \parP_t{\TP^{ab...}}_{\quad \ \ cd...}$ does not hold in general. It only holds if the $\xP^a$ coordinates do not depend on time. Instead we have the relation
\begin{align}
	\left({J_i}^a{J_j}^b \, ... \right) \pargal_t\left(\tensor{\Tgal}{^i^j^{...}_k_l_{...}}\right) &\left({J^k}_c{J^l}_d \, ...\right)  = \label{eq::par_t_gen} \\
	&\parP_t \tensor{\TP}{^a^b^{...}_c_d_{...}} - \Lie{\T \U}\tensor{\TP}{^a^b^{...}_c_d_{...}} \nonumber ,
\end{align}
where $\T \U$ is the coordinate velocity vector of the $\xP^a$ coordinates with respect to the $\xgal^i$ coordinates and is defined such as
\begin{align}
	\Ugal^i := \parP_t\xgal^i, \label{eq::def_UU}
\end{align}
which implies $\UP^a = -\pargal_t\xP^a$ using~\eqref{eq::par_t_gen}.

\proof{For simplicity we show the proof for a rank-1 tensor; it can easily be generalised for any tensor. Making use of $\pargal_t = \parP_t - \parP_t\xgal^k\pargal_k$, we have
\begin{align*}
	{J^i}_a\pargal_t \Tgal_i	&= {J^i}_a\left(\parP_t\convG T_i - \parP_t \xgal^k\pargal_k\convG T_i\right) \\
							&= \parP_t\TP_a - \convG T_i\parP_t{J^i}_a - {J^i}_a\Ugal^k\pargal_k\convG T_i \\
							&= \parP_t\TP_a - {J^i}_a\left(\convG T_k \, \pargal_i\Ugal^k + \Ugal^k\pargal_k\convG T_i\right) \\
							&= \parP_t\TP_a - {J^i}_a\left(\Lie{\T \U}\Tb T\right)_i \\
							&= \parP_t\TP_a - \Lie{\T \U} \TP_a. & \square
\end{align*}} \\

We can then write the system~\eqref{eq::Cont_New}-\eqref{eq::g_def} in the coordinates $\xP^a$. This gives \textit{the generalised Newton equations for the fluid velocity vector $\T \vN$} in a time parametrised coordinate system:
\begin{flalign}
	&\left(\parP_t + \Lie{\T \vN - \T \U}\right) \rho		= - \rho \theta,& \label{eq::Cont_NewP} \\
	&\left(\parP_t + \Lie{\T \vN - \T \U}\right) \theta		= -4\pi G \rho +\Lambda - \convP\Theta_{cd}\convP\Theta^{cd} + \convP\omega_{cd}\convP\omega^{cd}, \label{eq::Ray_NewP}& \\
	&\left(\parP_t + \Lie{\T \vN - \T \U}\right)\convP\omega_{cd}	= 0,& \label{eq::Wg_NewP}
\end{flalign}
and the definition of the gravitational field
\begin{align}
	\gP^a := \left(\parP_t + \Lie{\T \vN - \T \U}\right)\vNP^a + \vNP^c\left({\convP\Theta_c}^{\ a} + {\convP\omega_c}^{\ a}\right). \label{eq::g_defP}
\end{align}

These equations, while written for any parametrised coordinate system, still require reference coordinates, i.e. the fixed coordinates $\xgal^i$, to be able to define the tensor $\T \U$. This is discussed in Sec.~\ref{sec::Euler-Newton_discu}.

The system~\eqref{eq::Cont_NewP}-\eqref{eq::g_defP} depends on the vectors $\T \U$ and $\T \vN$. The latter can be called the velocity of the fluid with respect to the fixed coordinates. However when taking a non-zero coordinate velocity $\T \U$, it might be useful to work with the velocity vector $\T \V$ of the fluid with respect to the parametrised coordinates defined as
\begin{align}
	\T \V := \T \vN - \T \U.
\end{align}
In the next section we develop the Newton equations as functions of $\T \U$ and $\T \V$.

\subsubsection{In general parametrised coordinates -- $\T \V$ description}
\label{sec::Euler-Newton_PV}

We introduce the expansion tensors $\tensor[^{\V}]{\T{\Theta}}{}$ and $\tensor[^{\U}]{\T{\Theta}}{}$, and the vorticity tensors $\tensor[^{\V}]{\T{\omega}}{}$ and $\tensor[^{\U}]{\T{\omega}}{}$ of the vectors $\T \V$ and $\T \U$ as
\begin{align}
	\tensor[^\V]{\convP\Theta}{_a_b} &:= \DP_{(a} \VP_{b)} \quad ; \quad \tensor[^{\U}]{\convP\Theta}{_a_b} := \DP_{(a} \UP_{b)} \quad ; \nonumber \\ \tensor[^\V]{\convP\omega}{_a_b} &:= \DP_{[a}\VP_{b]} \quad ; \quad \tensor[^\U]{\convP\omega}{_a_b} := \DP_{[a}\UP_{b]}, \nonumber
\end{align}
and their trace $\tensor[^\V]{\theta}{} := \tensor[^\V]{\convP\Theta}{_c^c}$ and $\tensor[^\U]{\theta}{} := \tensor[^\U]{\convP\Theta}{_c^c}$. \\

We can then write the system~\eqref{eq::Cont_NewP}-\eqref{eq::g_defP} as function of $\T \V$ and $\T \U$. This gives \textit{the generalised Newton equations for the fluid coordinates velocity vector} $\T \V$ in a time parametrised coordinate system:
\begin{flalign}
	&\left(\parP_t + \Lie{\T \V}\right) \rho			= - \rho \left(\tensor[^\V]{\theta}{} + \tensor[^\U]{\theta}{}\right), \label{eq::Cont_NewVU} \\
	&\left(\parP_t + \Lie{\T \V}\right) \left(\tensor[^\V]{\theta}{} + \tensor[^\U]{\theta}{}\right)		= -4\pi G \rho + \Lambda \label{eq::Ray_NewVU} \\
		& \hspace{3.cm} - \left(\tensor[^\V]{\convP\Theta}{_c_d} + \tensor[^\U]{\convP\Theta}{_c_d}\right)\left(\tensor[^\V]{\convP\Theta}{^c^d} + \tensor[^\U]{\convP\Theta}{^c^d}\right) \nonumber \\
		& \hspace{3.cm} + \left(\tensor[^\V]{\convP\omega}{_c_d} + \tensor[^\U]{\convP\omega}{_c_d}\right)\left(\tensor[^\V]{\convP\omega}{^c^d} + \tensor[^\U]{\convP\omega}{^c^d}\right),\nonumber \\
	&\left(\parP_t + \Lie{\T \V}\right)\left(\tensor[^\V]{\convP\omega}{_c_d} + \tensor[^\U]{\convP\omega}{_c_d}\right)	= 0, \label{eq::Wg_NewVU}
\end{flalign}
and the definition of the gravitational field
\begin{align}
	\gP^a := &\left(\parP_t + \Lie{\T \V}\right)\left(\VP^a + \UP^a\right) \label{eq::g_defVU} \\
		&+ \left(\VP^c + \UP^c\right)\left(\tensor[^\V]{\convP\Theta}{_c^a} + \tensor[^\U]{\convP\Theta}{_c^a} + \tensor[^\V]{\convP\omega}{_c^a} + \tensor[^\U]{\convP\omega}{_c^a}\right). \nonumber
\end{align}

\subsubsection{Class of coordinates}
\label{sec::Euler-Newton_discu}

We define the following mathematical object: \\

\definition{Given a coordinate system $\{y^a\}_{a=1,2,3}$, parametrised or not, we define \textbf{the class of $y$ coordinates}, denoted $\classN{y}$, as the ensemble of coordinate systems which can be obtained from the system  $\{y^a\}_{a=1,2,3}$ with a time-independent coordinate transformation.}

The equations developed in Secs.~\ref{sec::Euler-Newton_fixed}, \ref{sec::Euler-Newton_Pv} and \ref{sec::Euler-Newton_PV} were defined, directly or indirectly, with respect to a chosen fixed coordinate system $\{\xgal^i\}_{i=1,2,3}$ on $\SNew$. They however do not depend on this system as all these equations are invariant under a time-independent change of coordinates. Instead, any system of coordinates in the class of fixed coordinates can be chosen. The same applies for the definition of the vector $\T \U$, and so of the vector $\T \V$. \\

\proof{We consider two fixed coordinate systems $\{\xgal^i\}_{i=1,2,3}$ and $\{\ygal^I\}_{I=1,2,3}$ and a parametrised coordinate system $\{\xP^a\}_{a=1,2,3}$. The components of a tensor in the $\ygal^I$ coordinates will be denoted with capital Roman letters $I$, $J$, ... Let $\T \U$ be the coordinate velocity vector of the coordinates $\xP^a$ with respect to the coordinates $\xgal^i$. Then
\begin{align*}
	\Ugal^i	&:= \parP_t\xgal^i \\
			&= \parP_t\xgal^i(\ygal^I) \\
			&= \pargal_t \ygal^K \ \partial_{\ygal^K}x^i.
\end{align*}
$\partial_{\ygal^K}x^i$ is the Jacobian of the coordinate transformation between $\xgal^i$ and $\ygal^K$. Then $\Ugal^I := \parP_t\ygal^I$. This means that the definition of $\T \U$ is unchanged if the fixed system of reference is $\{\ygal^I\}_{I=1,2,3}$. \hfill $\square$} \\

The choice of parametrised coordinates $\xP^a$ then defines uniquely the vector $\T \U$. The opposite is wrong: defining a vector field $\T \U$ on $\SNew$ does not determine uniquely a parametrised system $\{\xP^a\}_{a=1,2,3}$. Instead $\T \U$ uniquely defines a \textbf{class of parametrised coordinate systems} which we can write $\classN{\T \U}$. Then $\T \U$ is the coordinate velocity of any system in $\classN{\T \U}$ with respect to any system in $\classN{0}$, where $\classN{0}$ is the class of fixed coordinates.

The Newton system~\eqref{eq::Cont_NewP}-\eqref{eq::g_defP}, or equivalently the system~\eqref{eq::Cont_NewVU}-\eqref{eq::g_defVU}, then corresponds to the original Newton system~\eqref{eq::Cont_New}-\eqref{eq::g_def} written in any class of coordinates. It is the most general writing of the original equations~\eqref{eq::Cont_New}-\eqref{eq::g_def}, assuming the time parameter is unchanged.\footnote{Making a change of parametrisation $t \rightarrow \tilde t$ corresponds to a change of foliation for the 4-dimensional Newton equations (see Sec.~\ref{sec::3+1-Newton}).}

But while the original set of equations required the definition of only one vector field, the fluid velocity vector $\T \vN$, the general equations of Sec.~\ref{sec::Euler-Newton_Pv} require the definition of a second vector field, the coordinate velocity vector $\T \U$ of the chosen class of coordinates to work in. If one chooses the point of view of Sec.~\ref{sec::Euler-Newton_PV}, the pair of vectors $(\T \vN, \T \U)$ is replaced by the pair $(\T \V, \T \U)$. However only $\T \vN$ is  physical as it is the fluid velocity vector and does not depend on a chosen class of coordinates: taking $\T \vN = 0$ changes the generality of the equations as it implies $\pargal_t\rho = 0$, while $\T \U$ or $\T \V$ can be taken to $0$ without loss of generality.

A non-trivial choice of $\T \U$ can however be of physical interest depending on the physical system studied. In the next section we present specific examples of parametrised coordinates.

\subsection{Specific choices of coordinates}

In this section we will always use the Newton equations in the same class of coordinates as the vector $\T \U$ we will choose. We can then omit the tilde notation. The partial time derivative will also always be partial time derivative at fixed $\mathcal{X}_{\T \U}$ coordinates, we will note it $\derivtN{\T\U}$\footnote{To avoid possible confusions, we precise that this notation does not imply $\derivtN{\T\U} U^a = 0$.}.

\subsubsection{Galilean coordinates}

\textit{Galilean coordinates} are the classes of coordinates for which $\derivtN{\T\U} \U^a = 0$ and $\T \nabgal \T U = 0$, i.e. the coordinates $\mathcal{X}_{\T \U}$ are uniformly moving with respect to the class of fixed coordinates. If one chooses the fluid description in terms of $(\T \vN, \T \U)$, then the corresponding Newton system~\eqref{eq::Cont_NewP}-\eqref{eq::g_defP} is not equivalent for all Galilean coordinates due to the terms $\U^c\D_c$. The Galilean invariance only appears in the $(\T \V, \T \U)$ description of the fluid as the corresponding Eqs.~\eqref{eq::Cont_NewVU}-\eqref{eq::g_defVU} are formally equivalent for all Galilean coordinates.

This shows that the description in terms of the fluid coordinate velocity is more appropriate when $\T \U$ is non-zero as it will encode the non-inertial effect due to $\T \U$. Indeed, we can rewrite the Euler equation~\eqref{eq::g_defVU} to feature the non-inertial terms acting on $\T \V$
 \begin{align}
	 \left(\derivtN{\T\U} + \V^c\D_c\right)\V^a =	\ &g^a - \left(\derivtN{\T\U} + \U^c\D_c\right)\U^a \label{eq::g_defVU_bis} \\
	 								&- 2\V^c\left( \tensor[^\U]{\theta}{_c^a} + \tensor[^\U]{\convP\omega}{_c^a}\right). \nonumber
\end{align}
We see that the acceleration of $\T \V$, on the left-hand side of the equation, is affected by the gravitational field and the non-inertial terms, depending on $\T \U$. We however recall that these effects are only gauge effects as the true dynamics of the fluid is given by $\T \vN$.

\subsubsection{Globally translating and rotating coordinates}
\label{sec::glob_trans_rot}

In classical mechanics, the most general coordinates are usually globally rotating and translating with respect to $\mathcal{X}_0$. In this case they are called \textit{frames}. They correspond to all the classes of coordinates where $\T \U$ can be decomposed as
\begin{align}
	\T \U = \T \U_{\rm tr} + \T \U_{\rm rot},
\end{align}
where $\T \D \T \U_{\rm tr} = 0$, $\tensor[^\U]{\T\Theta}{_{\rm rot}} = 0$ and $\epsilon_{acd}\D_b\D^c\U^d_{\rm rot} = 0$ where $\T\epsilon$ is the Levi-Civita tensor.

The condition $\T \D \T \U_{\rm tr} = 0$ ensures that $\T \U_{\rm tr}$ is a global translation of the $\mathcal{X}_{\T \U}$ coordinates with respect to the Galilean classes of coordinates; $\tensor[^\U]{\T\Theta}{_{\rm rot}} = 0$ ensures that $\T \U_{\rm rot}$ is only rotational. The rotation vector of the frame is $\Omega ^a := {\epsilon^a}_{cd}\D^c\U^d_{\rm rot}$. Then $\epsilon_{bcd}\D_a\D^c\U^d_{\rm rot} = \D_a\Omega_b = 0$ ensures that this rotation is also global.

In these conditions, the Euler equation~\eqref{eq::g_defVU_bis} becomes
 \begin{align}
	 \left(\derivtN{\T\U} + \V^c\D_c\right)\V^a = \	& g^a - \derivtN{\T\U} \U_{\rm tr}^a - \derivtN{\T\U} \U_{\rm rot}^a \label{eq::g_defVU_global} \\
	 & - \left(\U^c_{\rm tr} + \U^c_{\rm rot}\right){\omega_c}^{a} - 2\V^c \, {^\mathrm{\U}{\omega_c}}^{a}. \nonumber
\end{align}
The term $ \derivtN{\T\U}\U_{\rm tr}^a + \derivtN{\T\U}\U_{\rm rot}^a + \left(\U^c_{\rm tr} + \U^c_{\rm rot}\right){\omega_c}^{a}$ is the \textit{centrifugal acceleration}, and $2\V^c \, {^\mathrm{\U}{\omega_c}}^{\ a}$ is the \textit{Coriolis acceleration}. We retrieve the usual Euler equation in a non-inertial frame where the vorticity of $\T \U$ corresponds to the global rotation of that frame with respect to a Galilean frame. There is however no contribution of the expansion tensor of $\T \U$ in that case, as it is zero. In the next section we will show to what corresponds a non-zero $\tensor[^\U]{\T\Theta}{}$.

\subsubsection{Homogeneous deformation}
\label{sec::Hubble_flow_New}

The expansion tensor of the coordinate velocity vector $\T \U$ can be linked to the time variations of the metric in the $\mathcal{X}_{\T \U}$ coordinates. We have the following relation:
\begin{align}
	\frac{1}{2}\derivtN{\T\U} h_{ab} = \tensor[^\U]{\Theta}{_a_b}. \label{eq::part_h}
\end{align}
\proof{For this proof only we reintroduce the tilde and untilde notations of Sec.~\ref{sec::Euler-Newton_Pv} concerning parametrised and fixed coordinates. Using property~\eqref{eq::par_t_gen}, we have
\begin{align*}
	{J^i}_{a}{J^j}_{b}\derivtN{0} \hgal_{ij}	&= \left(\derivtN{\T\U} - \Lie{\T \U}\right) \hP_{ab}, \\
								&= \derivtN{\T\U} \hP_{ab} - 2\DP_{(a}\UP_{b)}.
\end{align*}
$\derivtN{0}$ is the time derivative with respect to the fixed coordinate class. Because $\xgal^i$ are fixed coordinates, $\derivtN{0} \hgal_{ij} = 0$. Then $\derivtN{\T\U} \hP_{ab} - 2\DP_{(a}\UP_{b)} = 0$. \hfill $\square$} \\

\remark{In a frame coordinate system, i.e. globally translating and rotating, the metric is static as ${^\mathrm{\U}\T\Theta} = 0$.}

Relation~\eqref{eq::part_h} implies that with a change of coordinates from fixed coordinates, we can simulate space expansion. This expansion is always a \textit{gradient expansion}, i.e. $\frac{1}{2}\derivtN{\T\U} h_{ab}$ is the gradient of a vector. Taking $\tensor[^\U]{\T\Theta}{}$ such that $\T \D \, \tensor[^\U]{\T\Theta}{} = 0$ implies that the expansion is global: this is called a $\textit{homogeneous deformation}$. Furthermore, when it is isotropic, the coordinate velocity vector corresponds to the \textit{position vector}, i.e. in Cartesian coordinates $\U^a \propto x^a$. In this case this is called a \textit{Hubble flow}. However one has to remember that the physical vector is $\T \vN$. Therefore the expansion due to $\tensor[^\U]{\T\Theta}{}$ is strictly speaking a fluid expansion and not a space expansion (see Sec.~\ref{sec::hyp} for precisions on this interpretation). \\

The main consequence of the gradient expansion is that no global expansion is possible if the 3D-manifold $\SNew$ has a compact topology\footnote{The only possible compact oriented topology is the flat 3-torus~$\mathbb T^3$ up to a finite covering.}. Indeed  in such a topology $x^a$ cannot be the components of a tensor as they do not respect the global symmetry of a compact space. So strictly speaking, Newtonian cosmological simulations, said to be realised in a 3-torus with global isotropic expansion, are actually simulating an infinite 3D-manifold. The 3-torus symmetry is only set on $\T \V$ and not $\T \U$, thus the physical vector $\T \vN$ lies in an infinite 3D-manifold. If one wants  $\SNew$ to be strictly compact and be allowed for expansion, modified-Newtonian equations have to be used. This is discussed in Sec.~\ref{sec::Post_New_exp}.

\subsubsection{Lagrangian coordinates}
\label{sec::Lag_New}

We saw that the physical dynamical properties of the fluid are encoded in $\T \vN$. By working in parametrised coordinates, we split these properties into $\T \U$ and $\T \V$. Then, taking $\T \V = 0$ implies that the coordinate velocity $\T \U$ is the velocity of the fluid $\T\vN$. Coordinates such as $ \T \V = 0$ are called \textit{Lagrangian coordinates} as they follow the fluid flows given by $\T \vN$. In Lagrangian coordinates, part of the fluid dynamics, the pure expansion $\theta$ and the shear $\sigma_{ab} := \Theta_{\langle ab\rangle}$ of $\T \vN$, is put into the time variation of the metric. The other part, the vorticity of $\T \vN$, does not affect the metric.

\subsection{Similarities with the 3+1 and 1+3 formalisms of general relativity}

In Sec.~\ref{sec::Euler-Newton}, we derived the Newton system of equations in an arbitrary class of coordinates. We saw that the freedom associated with a choice of class is a vector $\T\U$. Furthermore the difference between two partial time derivatives is a Lie derivative. These two properties resemble the properties of the shift freedom in the 3+1 formalism of general relativity (see Sec.~\ref{sec::3+1_GR}). We could add that Newton's equations live on a time-parametrised 3D-manifold\footnote{Actually, among the tensors defining $\SNew$, only the metric can depend on time, the Riemann tensor being zero in any class of coordinates.} which is the same situation as for the 3+1-Einstein equations.

This shows that apart from the known formal equivalence between the Newtonian system~\eqref{eq::Cont_New}-\eqref{eq::Wg_New} and the 1+3-Einstein equations explained in Ref.~\cite{1971_Ellis}, Newton also features similarities with the 3+1 construction of the Einstein equations. In the next section, we will reverse this construction in the case of the Newton theory to get 4D-Newton equations.

\section{The 4D-Newton system} 
\label{sec::4D_Newt}

We recall in Sec.~\ref{sec::3+1_GR} the construction of the 3+1 equations of GR. Reversing this construction will allow us to write the 4D-Newton equations in Sec.~\ref{sec::4D_Newton}. We also quickly present the 1+3-Einstein equations in Sec.~\ref{sec::1+3-Einstein} as they will be formally equivalent to the 4D-Newton system for a certain choice of manifold (see Sec.~\ref{sec::Ortho}).

\subsection{3+1 formalism in general relativity}
\label{sec::3+1_GR}

We define a 4D pseudo-Riemannian manifold $\CM$, called the spacetime manifold, and its metric $\T g$. This metric has a Lorentzian signature $(-+++)$. In the following subsections we will derive, from the Einstein equations, the 3+1-Einstein equations on a 3-dimensional manifold.

\subsubsection{Foliation variables}

The principle behind the 3+1 formalism is to split the spacetime manifold $\CM$ into space and time. If $\CM$ is globally hyperbolic, which we will suppose from now, it is possible to define a family of spacelike hypersurfaces $\folGR$ in $\CM$. This family is called a \textit{foliation} and can be uniquely defined by the level surfaces of a smooth scalar field $\hat{t}$ on $\CM$.

The 3+1-Einstein equations are the projections of the Einstein equation onto and normal to the foliation $\folGR$. To be able to realise these projections one has to define a normal unit vector field to the family of hypersurfaces. The gradient $\T\nabla \hat{t}$ of the scalar field $\hat{t}$ defines naturally a normal timelike vector field to the hypersurfaces. In general this vector is not a unit vector. We then define the timelike unit vector field to the hypersurfaces $\Sigma_t$ as
\begin{align}
	\T n := - N\T\nabla \hat{t},
\end{align}
where $N := \left(- \nabla_\mu \hat{t} \, \nabla^\mu \hat{t}\right)^{-1/2}$ is called the \textit{lapse}, is positive by convention, and only depends on the foliation. The global minus sign in the definition of $\T n$ is a convention imposing this vector to be future oriented with respect to the time scalar field $\hat{t}$. The 3+1-Einstein equations we will get do not depend on this convention.

The projection operator on to the hypersurfaces is the tensor
\begin{align}
	\T h := \T g + \T n \otimes \T n,
\end{align}
where $\T g$ is the metric on $\CM$.

A \textit{spatial tensor} is defined as having no normal part with respect to the hypersurfaces $\Sigma_t$. The spatial covariant derivative $\T D$ applied on a spatial vector $\T T$ is defined as
\begin{align}
	D_\mu {T^{\alpha_1 ...}}_{\beta_1 ...} := {h^\sigma}_\mu \left({h^{\alpha_1}}_{\mu_1} ... \right) \left({h^{\nu_1}}_{\beta_1} ... \right) \nabla_\sigma {T^{\mu_1 ...}}_{\nu_1 ...}. \label{eq::D_def}
\end{align}

We define two more spatial rank-2 tensors, the intrinsic Ricci curvature $\T R$ of the hypersurfaces $\Sigma_t$ and the extrinsic curvature $\T K$ of these hypersurfaces embedded in $\CM$. The extrinsic curvature makes the link between the geometrical properties of the hypersurfaces $\Sigma_t$ and the ones of $\CM$. We can write the components of $\T K$ as
\begin{align}
	K_{\alpha\beta} = - {h^\mu}_\alpha {h^\nu}_\beta \nabla_\nu n_\mu, \label{eq::K_def_1}
\end{align}
or
\begin{align}
	K_{\alpha\beta} = - \frac{1}{2N}\Lie{N\T n}h_{\alpha\beta}. \label{eq::K_def_2}
\end{align}
The negative sign is a convention. Because $\T n$ is proportional to a gradient, $\T K$ is a symmetric tensor. Then the gradient of the normal vector can be decomposed as
\begin{align}
	\nabla_\alpha n_\beta = -K_{\beta\alpha} - n_\alpha \tensor[^n]{a}{_\beta},
\end{align}
where $\tensor[^n]{a}{_\beta}$ is the 4-acceleration of the normal vector with $\tensor[^n]{a}{_\alpha} := n^\mu\nabla_\mu n_\alpha = \D_\alpha \ln N$.

\subsubsection{3+1 decomposition of the spacetime Ricci tensor}

We give in the present section the decomposition of the Ricci curvature tensor ${\tensor[^4]{\T R}{}}$ of $\CM$ onto the foliation and orthogonal to it. ${\tensor[^4]{\T R}{}}$ being a symmetric tensor, we will have 10 projection equations. Details for the derivation of these equations can be found in Ref.~\cite{2012_GG}.

The two times projection onto $\Sigma_t$ gives the 3+1-Ricci\footnote{We name the equations with the suffix "3+1" to distinguish them from their equivalent in the 1+3 formalism of general relativity (see Ref.~\cite{2014_Roy}).} equation
\begin{align}
	{h^\mu}_\alpha {h^\nu}_{\beta} {\Riem_{\mu\nu}} = &-\frac{1}{N}\Lie{N\T n}K_{\alpha\beta} - \frac{1}{N} \D_\alpha\D_\beta N \nonumber \\
	&+ R_{\alpha\beta} + K K_{\alpha\beta} - 2K_{\alpha\mu}{K^\mu}_\beta, \label{eq::Riem_Ricci}
\end{align}
where $K$ is the trace of $\T K$. Note that this equation features only spatial tensors as the Lie derivatives of a spatial tensor along $N \T n$ (or along $\T n$) is a spatial tensor.

The spatial and orthogonal projection gives the 3+1-Codazzi equation
\begin{align}
	{h^\mu}_\alpha n^\nu {\Riem_{\mu\nu}} = \D_\alpha K - \D_\mu{K^\mu}_\alpha.  \label{eq::Riem_Codazzi}
\end{align}

The two times orthogonal projection gives the 3+1-Raychaudhuri equation
\begin{align}
	n^\mu n^\nu\Riem_{\mu\nu} = &\frac{1}{N}\Lie{N\T n} K - K_{\mu\nu}K^{\mu\nu} \nonumber \\
	&+ \frac{1}{N}\D_\mu \D^\mu N. \label{eq::Riem_Ray}
\end{align}

Combining the trace of the 3+1-Ricci equation~\eqref{eq::Riem_Ricci} with the 3+1-Raychaudhuri equation~\eqref{eq::Riem_Ray} we obtain the 3+1-Gauss equation
\begin{align}
	\Riem + 2 \,{\Riem_{\mu\nu}}n^\mu n^\nu = R + K^2 - K_{\mu\nu}K^{\mu\nu}.  \label{eq::Riem_Gauss}
\end{align}
Note that 3+1-Gauss is redundant with the 3+1-Ricci and 3+1-Raychaudhuri equations together, it is however essential when solving the Cauchy problem in general relativity.

\subsubsection{3+1-Einstein equations}
\label{sec::3+1-Einstein}

We consider now that $\CM$ is solution to the Einstein equations. Then ${\tensor[^4]{\T R}{}}$ is solution of
\begin{align}
	\Riem_{\alpha\beta} - \frac{\Riem}{2} g_{\alpha\beta} + \Lambda g_{\alpha\beta} = 8\pi G T_{\alpha\beta},
\end{align}
where $\T T$ is the stress-energy tensor of the matter in $\CM$ and can be decomposed with respect to the foliation $\folGR$ as
\begin{align}
	T_{\alpha\beta} &= E \, n_\alpha n_\beta + P h_{\alpha\beta} + 2Q_{(\alpha}n_{\beta)} + \Pi_{\alpha\beta}. \label{eq::T^n}
\end{align}
$E$, $P$, $\T Q$, $\T \Pi$ are respectively the energy density, the pressure, the heat flux, and the anisotropic stress of the matter as measured by an observer of 4-velocity $\T n$. We call such an observer an \textit{Eulerian observer}. By definition $Q_\mu n^\mu = 0$ and $\Pi_{\alpha\beta} = \Pi_{(\alpha\beta)}$ with $\Pi_{\alpha\mu} n^\mu = 0$ and ${\Pi_{\mu}}^\mu =0$.

The 3+1-Einstein system of equations is obtained from Eqs.~\eqref{eq::Riem_Ricci}-\eqref{eq::Riem_Gauss} when introducing the previous matter variables. Written as a Cauchy system it is composed of 6 evolution equations, obtained from the 3+1-Ricci equation,
\begin{align}
	\frac{1}{N}&\Lie{N\T n} K_{\alpha\beta} = 4\pi G \left[-\left(E - P\right) h_{\alpha\beta} + 2 \Pi_{\alpha\beta}\right] \label{eq::T_Ricci} \\
									&  - \Lambda \, h_{ab} - \frac{1}{N} \D_\alpha\D_\beta N + R_{\alpha\beta} + K K_{\alpha\beta} - 2K_{\alpha\mu}{K^\mu}_\beta, \nonumber
\end{align}
and two constraint equations, the momentum constraint~\eqref{eq::T_Codazzi} (or 3+1-Codazzi equation) and the Hamilton constraint~\eqref{eq::T_Gauss} (or 3+1-Gauss equation):
\begin{align}
	- 8\pi G Q_\alpha &= \D_\alpha K - \D_\mu{K^\mu}_\alpha,  \label{eq::T_Codazzi} \\
	16\pi G E + 2 \Lambda &= R + K^2 - K_{\mu\nu}K^{\mu\nu}.  \label{eq::T_Gauss}
\end{align}
Like in the previous section, combining the 3+1-Ricci equation~\eqref{eq::T_Ricci} and the Hamilton constraint~\eqref{eq::T_Gauss}, we get the 3+1-Raychaudhuri equation
\begin{align}
	\frac{1}{N}\Lie{N\T n} K = &- 4\pi G\left(E + 3P\right) + \Lambda + K_{\mu\nu}K^{\mu\nu} \nonumber \\
		&- \frac{1}{N}\D_\mu \D^\mu N. \label{eq::T_Ray}
\end{align}
While this last equation is not part of the Cauchy problem of the 3+1-Einstein equations, we keep it as it will be useful for further comparisons with Newton.

\subsubsection{3+1-conservation equations}
\label{sec::3+1-conservation}

Solving the set of Eqs.~\eqref{eq::T_Ricci}-\eqref{eq::T_Gauss} is sufficient to solve the Einstein equations. It is however of physical relevance to give two additional equations, that is the \textit{3+1-energy conservation} and the \textit{3+1-momentum conservation} both coming from the projections of the conservation equation $\nabla_\mu \tensor{T}{^\mu_\alpha} = 0$ with respect to $\folGR$.

The 3+1-energy conservation is
\begin{align}
	\frac{1}{N}\Lie{N\T n} E = \ & K \left(E+P\right) - \D_\mu Q^\mu - 2Q^\mu\D_\mu\ln N \nonumber \\
	&+  K_{\mu\nu} \Pi^{\mu\nu}, \label{eq::3+1-Energy}
\end{align}
and the 3+1-momentum conservation is
\begin{align}
	\frac{1}{N}\Lie{N\T n} Q_\alpha = &- \left(E + P\right)\D_\alpha \ln N - \D_\alpha P + K Q_\alpha \nonumber \\
	&- \D_\mu\tensor{\Pi}{^\mu_\alpha} - \tensor{\Pi}{^\mu_\alpha} \D_\mu\ln N. \label{eq::3+1-Momentum}
\end{align}

\subsubsection{The matter fluid}
\label{sec::Tmunu}

We assume the matter is a fluid of 4-velocity $\T u$ with $u_\mu u^\mu = -1$, i.e. a non-radiative fluid. \\

\paragraph{Stress-energy tensor.} The stress-energy tensor can be decomposed with respect to the fluid 4-velocity as
\begin{align}
	T_{\alpha\beta} &= \epsilon u_\alpha u_\beta + p b_{\alpha\beta} + 2q_{(\alpha}u_{\beta)} + \pi_{\alpha\beta}, \label{eq::T^u}
\end{align}
where $\epsilon$ is the energy density, $p$ the pressure, $\T q$ the heat flux and $\T \pi$ the anisotropic stress of the fluid as measured in its rest frames. $\T b$ is the projector on the rest frames of the fluid, with $\T b := \T g + \T u \otimes \T u$. As for $\T Q$ and $\T \Pi$, by definition, $q_\mu u^\mu = 0$ and $\pi_{\alpha\beta} = \pi_{(\alpha\beta)}$ with $\pi_{\alpha\mu} u^\mu = 0$ and ${\pi_{\mu}}^\mu =0$. 

For a general foliation, $\T n \not= \T u$. Then the fluid variables measured in the rest frames are different from the one measured by the Eulerian observer. The nature of the fluid is however given by the variables measured in the rest frames. It is therefore often useful to express the variables measured by $\T n$ [defined in~\eqref{eq::T^n}] as function of the ones measured by $\T u$ [defined in~\eqref{eq::T^u}]. For this we introduce the \textit{tilt velocity} $\T \vt$ of the fluid 4-velocity with respect to the foliation as
\begin{align}
	\T \vt := \frac{1}{\gamma}\T u - \T n, \label{eq::tilt_vel}
\end{align}
where $\gamma := \left(1 - \vt_\mu \vt^\mu\right)^{-1/2}$ is the Lorentz factor. $\T \vt$ is spatial by definition.

We then have the following relations:
\begin{align}
	E			= \ & \gamma^2\epsilon + \left(\gamma^2-1\right)p + 2\gamma \vt^\mu q_\mu + \vt^\mu \vt^\nu\pi_{\mu\nu}, \\
	P			= \ & \left(\gamma^2-1\right)\epsilon + \left(\gamma^2+2\right)p + 2\gamma \vt^\mu q_\mu + \vt^\mu \vt^\nu\pi_{\mu\nu}, \\
	Q_\alpha		= \ & \gamma^2\left(\epsilon+p\right)\vt_\alpha + \gamma \vt^\mu q_\mu \vt_\alpha + \gamma{h^\mu}_\alpha q_\mu  - {h^\mu}_\alpha \vt^\nu \pi_{\mu\nu}, \\
	\Pi_{\alpha\beta} = \ & \gamma^2\epsilon \, \vt_\alpha \vt_\beta + p\left(h_{\alpha\beta} + \gamma^2 \vt_\alpha \vt_\beta\right) + 2\gamma \vt_{(\alpha}{h_{\beta)}}^\mu q_\mu \nonumber \\
				&+ {h_{\alpha}}^\mu {h_{\beta}}^\nu \pi_{\mu\nu} - P h_{\alpha\beta},
\end{align}

We also give the specific example of a dust fluid, characterised by $q_\mu = 0 = \pi_{\mu\nu}$ and $p=0$:
\begin{align}
	E^{\rm(DF)}			&= \gamma^2\epsilon, \label{eq::TT_E} \\
	P^{\rm(DF)}			&= \frac{\gamma^2-1}{3}\epsilon, \label{eq::TT_P} \\
	Q^{\rm(DF)}_\mu		&= \gamma^2\epsilon \, \vt_\mu, \label{eq::TT_Q} \\
	\Pi^{\rm(DF)}_{\mu\nu}	&= \epsilon \left(\gamma^2 \vt_\mu \vt_\nu - \frac{\gamma^2-1}{3}h_{\mu\nu}\right), \label{eq::TT_Pi}
\end{align}
where we introduced the upper-script $^{\rm(DF)}$ to denote the Eulerian fluid variables for a dust fluid. \\

\paragraph{Kinematical variables.} As for Newton we define an expansion tensor $\T \Theta$ and a vorticity tensor $\T \omega$ of the fluid. They respectively correspond to the symmetric and antisymmetric part of the 4-velocity gradient $\T \nabla \T u$ projected on the rest frames of the fluid:
\begin{align}
	\Theta_{\alpha\beta}		&:= {b^\mu}_{(\alpha} {b^\nu}_{\beta)} \nabla_{\mu}u_{\nu}, \\
	\omega_{\alpha\beta}	&:={b^\mu}_{[\alpha} {b^\nu}_{\beta]} \nabla_{\mu}u_{\nu}. \label{eq::def_omega_GR}
\end{align}
Then the 4-velocity gradient can be decomposed as
\begin{align}
	\nabla_\alpha u_\beta = \Theta_{\alpha\beta} + \omega_{\alpha\beta} - u_{\alpha}\tensor[^u]{a}{_\beta}. \label{eq::grad_u}
\end{align}
where $\tensor[^u]{a}{^\alpha} := u^\mu\nabla_\mu u^\alpha$ is the 4-acceleration of the fluid. For a dust fluid $\tensor[^u]{\T a}{} = 0$ and we can rewrite the expansion and vorticity tensors as
\begin{align}
	\Theta_{\alpha\beta}^{\rm(DF)} = \nabla_{(\alpha} u_{\beta)}, \\
	\omega_{\alpha\beta}^{\rm(DF)} = \nabla_{[\alpha} u_{\beta]}.
\end{align}

In the case of a flow orthogonal foliation, i.e. $\T n~=~\T u$, we have the following relation between the extrinsic curvature and the expansion tensor, $\T K = - \T \Theta$. However, the vorticity in that case is necessarily zero because of the Frobenius theorem. So for vortical flows, if one wants to decompose the Einstein equation on a foliation, the fluid will necessarily be tilted with respect to that foliation.

It is however still possible to write the Einstein equation projected normal and orthogonal to the fluid. This gives the 1+3-Einstein equations (e.g. \cite{1971_Ellis,2014_Roy}) presented in Sec.~\ref{sec::1+3-Einstein}.

\subsubsection{Foliation adapted coordinates}
\label{sec::fol_adapt}

To formulate the Newton equations on a 4D-manifold we will need the mathematical tools used to write the 3+1-Einstein equations on a single 3D-manifold. This section aims at presenting these tools. \\

\paragraph{Shift vector and classes of adapted coordinates.}

The last tool we needed for construction of the 4D-Newton equations concerns the choice of coordinates. Until now we wrote the 3+1-Einstein equations for any coordinate system. We however often want to introduce one, and especially one which is adapted to the foliation. In such a coordinate system, the coordinate vector basis $\{\T{\partial}_\alpha\}_{\alpha=0,1,2,3}$ features three spatial vectors: $\T{\partial}_1$, $\T{\partial}_2$ and $\T{\partial}_3$. The 0-coordinate is chosen to correspond to the scalar field $\hat t$. We then write $\T{\partial}_t := \T{\partial}_0$ and call it the \textit{time vector}. By definition, $\T{\partial}_t$  is not spatial\footnote{This does not imply that $\T{\partial}_t$ is timelike (see section 5.2 in Ref.~\cite{2012_GG}). This will be discussed in relation with the 1+3-Newton equations in Sec.~\ref{sec::Ortho}.}.

In general, $\T{\partial}_t \not= \T n$, and we have ${\partial_t}^\mu n_\mu = - N$. We then define the \textit{shift vector} $\T \beta$ as
\begin{align}
	\T \beta := \T \partial_t - N \T n. \label{eq::def_beta}
\end{align}
By definition $\T \beta$ is spatial.

This vector plays the same role as $\T \U$ in Sec.~\ref{sec::Euler-Newton_discu}: instead of defining a single adapted coordinate system in $\CM$, it defines a class of coordinate systems adapted to the foliation $\folGR$ in $\CM$. We write this class $\classn{\T \beta}$. By definition, any coordinate system in $\classn{\T \beta}$ can be obtained from a coordinate system having $\T \beta$ as shift vector with a time independent spatial change of coordinates, i.e. $(t \rightarrow t, x^a \rightarrow y^b(x^a))_{a,b=1,2,3}$.

Reversing the definition~\eqref{eq::def_beta}: to a spatial vector $\T\beta$ corresponds a time vector
\begin{align*}
	\Tderivtn{\T\beta} := N\T n + \T\beta
\end{align*}
whose partial time derivative $\derivtn{\T\beta}$ is at fixed $\classn{\T \beta}$ coordinates.

The class with a zero shift, denoted $\classn{0}$ and its time vector $\Tderivtn{0}$, is said to be comoving with respect to the Eulerian observer. We call these coordinates \textit{Eulerian comoving coordinates}. Then any shift $\T \beta$ corresponds to the coordinate velocity vector of the class $\classn{\T\beta}$ with respect to these coordinates. \\

\paragraph{Pull-back.}

Once we chose an adapted coordinate system $\{t, x^a\}_{a=1,2,3}$, characterised by its shift $\T \beta$, it is possible to write the 3+1-Einstein equations~\eqref{eq::T_Ricci}-\eqref{eq::T_Ray} with indices running from 1 to 3. This comes from the fact that the 4D-components ${T^{\alpha_1 ...}}_{\beta_1...}$ of any spatial tensor $\T T$ are totally determined by the spatial components ${T^{a_1 ...}}_{b_1...}$ and by the spatial components $\beta^a$ of the shift (see Eq.~\eqref{eq::ind_red}). Note that the shift is only needed for covariant components.

For instance, in a class $\classn{\T \beta}$, the contravariant components of a rank-1 spatial tensor $\T A$ are $A^\alpha = (0, V^a)$, and its covariant components are $V_\alpha = (\beta^c V_c, V_a)$. The spatial covariant components can be obtained from the contravariant ones by lowering with the spatial components $h_{ab}$ of the spatial metric, i.e. $V_a = V^c h_{ac}$. The same can be done with a rank-2 tensor $\T T$:
\begin{align}
	T^{\alpha\beta} = \left(\begin{array}{c|c}0 & 0 \\\hline 0 & T^{ab}\end{array}\right) \ ; \ T_{\alpha\beta} = \left(\begin{array}{c|c}\beta^c\beta^d T_{cd} & \beta^cT_{ca} \\\hline T_{ac}\beta^c & T_{ab}\end{array}\right). \label{eq::ind_red}
\end{align}

The operation $T_{\alpha\beta} \rightarrow T_{ab}$ is called a \textit{pull-back}. It links spatial tensors on $\CM$ to tensors on a single 3-dimensional manifold $\Sigma$. The components of the pulled-back tensor $\T T$ on $\Sigma$ are $T_{ab}$. As for each hypersurface $\Sigma_t$ corresponds a pull-back to $\Sigma$, the global pull-back from $\CM$ to $\Sigma$ is said to be parametrised by time. This implies that the properties of the 3D-manifold $\Sigma$ and the tensors defined on it are parametrised by time. This situation is similar to the Newton theory in Sec.~\ref{sec::Newton}, where we had a time-parametrised 3D-manifold $\SNew$.

Applying the pulling-back operation on the 3+1-Einstein equations~\eqref{eq::T_Ricci}-\eqref{eq::T_Ray} allows us to have equations living on the 3D-manifold $\Sigma$ and parametrised by time. \\

\paragraph{3+1-Einstein equations on $\Sigma$.} To write the 3+1-equations on $\Sigma$ we need to pull-back each term of these equations. The only non-trivial term is one arising from the Lie derivative $\Lie{N\T n}$ present in the 3+1-Ricci~\eqref{eq::T_Ricci} and 3+1-Raychaudhuri~\eqref{eq::T_Ray} equations, as it still explicitly features a non-spatial tensor, i.e. $\T n$. To remove this dependence, we use the definition of the shift~\eqref{eq::def_beta} and the fact that $\Lie{\Tderivtn{\T\beta}} {T^{\alpha_1 ...}}_{\beta_1...} = \derivtn{\T\beta} {T^{\alpha_1 ...}}_{\beta_1...}$. Then for a spatial tensor $\T T$ of type $(n,m)$, the spatial components of $\Lie{N\T n}{\T T}$ are
\begin{align}
	{\left(\Lie{N\T n}{\T T}\right)^{a_1 ...}}_{b_1...} = \derivtn{\T\beta}  {T^{a_1 ...}}_{b_1...} - \tensor[^\Sigma]{{\Lie{\T \beta}}}{} {T^{a_1 ...}}_{b_1...}, \label{eq::Lie_ab}
\end{align}
where we introduced the notation $\tensor[^\Sigma]{{\Lie{}}}{}$ to denote the Lie derivative on $\Sigma$. According to the definition of the Lie derivative~\eqref{eq::Lie_def}, $\tensor[^\Sigma]{{\Lie{}}}{}$ uses the Levi-Civita connection on $\Sigma$, which corresponds to the pull-back of $\T \D$. \\

\remark{$\tensor[^\Sigma]{{\Lie{\T \beta}}}{} {T^{a_1 ...}}_{b_1...}$ corresponds to the spatial components of the spatial projection of $\tensor[]{{\Lie{\T \beta}}}{} {T^{\alpha_1 ...}}_{\beta_1...}$. The latter is however not necessarily spatial (see appendix~\ref{app::Lie}).}

Equations~\eqref{eq::ind_red} and~\eqref{eq::Lie_ab} allow us to write the 3+1-Einstein equations as equations living on the 3D-manifold $\Sigma$ parametrised by the time $t$. Then the 3+1-Ricci evolution equation becomes
\begin{align}
	\frac{1}{N}\Big( &\derivtn{\T\beta} - \tensor[^\Sigma]{{\Lie{\T \beta}}}{}\Big) K_{ab} = 4\pi G \left[-\left(E - P\right) h_{ab} + 2 \Pi_{ab}\right] \nonumber \\
									& - \Lambda \, h_{ab} - \frac{1}{N} \D_a\D_b N + R_{ab} + K K_{ab} - 2K_{ac}{K^c}_b; \label{eq::T_Ricci_3D}
\end{align}
the 3+1-constraints become
\begin{align}
	- 8\pi G Q_a			&= \D_a K - \D_c{K^c}_a,  \label{eq::T_Codazzi_3D} \\
	16\pi G E + 2\Lambda	&= R + K^2 - K_{cd}K^{cd};  \label{eq::T_Gauss_3D}
\end{align}
and the 3+1-Raychaudhuri equation becomes
\begin{align}
	\frac{1}{N}\left(\derivtn{\T\beta} - \tensor[^\Sigma]{{\Lie{\T \beta}}}{}\right) K = &- 4\pi G\left(E + 3P\right) + \Lambda + K_{cd}K^{cd} \nonumber \\
		&- \frac{1}{N}\D_c \D^c N. \label{eq::T_Ray_3D}
\end{align}

This concludes the construction of the 3+1-Einstein equations on a time parametrised 3D-manifold $\Sigma$ from the Einstein equation.

While detailing this construction, we saw  that the shift vector plays the same role as $\T\U$ in Sec.~\ref{sec::Euler-Newton_Pv}. The differential operator in the 3+1-Einstein equations is also similar to the one in the Newton equations~\eqref{eq::Cont_New}-\eqref{eq::Wg_New}. Using these similarities, we will be able to formulate the classical Newton equations as living in a 4D-manifold. This will be done in Sec.~\ref{sec::push_for} with the dual to the pull-back operation, i.e. a push-forward.

\subsection{1+3-Einstein equations}
\label{sec::1+3-Einstein}

We briefly present in this section the 1+3-Einstein system of equations (see Ref.~\cite{2014_Roy} for a more complete study). These equations correspond to the Einstein equation projected on $\T u$ and on the rest frames of the fluid, with the projector $\T b := \T g + \T u \otimes \T u$. The rest frames on which the equations are written however do not correspond to a family of hypersurfaces if the fluid is rotational.

For future comparisons with the 4D-Newton equations, we introduce two of the 1+3 equations, namely the \textit{1+3-Raychaudhuri equation} obtained by projecting twice ${\tensor[^4]{\T R}{}}$ on the fluid 4-velocity
\begin{align}
	\Lie{\T u}  \theta	= -4\pi G \epsilon - \Theta_{\mu\nu}\Theta^{\mu\nu} + \omega_{\mu\nu}\omega^{\mu\nu} + \nabla_\mu a^\mu, \label{eq::Ray_1+3} 
\end{align}
and the \textit{1+3-vorticity equation}
\begin{align}
	\Lie{\T u} \omega_{\alpha\beta} = {b^\mu}_{[\alpha}{b^\nu}_{\beta]} \nabla_\mu a_\nu. \label{eq::Wg_1+3}
\end{align}
This last equation is a geometrical constraint and does not require the Einstein equation to be valid.

Note that there also exists a 1+3-Ricci equation. It however requires to define a Riemann tensor on the rest frames of the fluid which does not have all the symmetries of the usual Riemann tensor (see Ref.~\cite{2014_Roy}). We do not introduce this equation here as it will not be useful for our discussion.

Finally the conservation equation $\nabla_\mu \tensor{T}{^\mu_\alpha}$ for the fluid stress-energy tensor gives the 1+3-energy conservation
\begin{align}
	\Lie{\T u} \epsilon &= - \theta \left(\epsilon+p\right) - \nabla_\mu q^\mu - \tensor[^u]{a}{_\mu} q^\mu -  \Theta_{\mu\nu} \pi^{\mu\nu}, \label{eq::1+3-Energy}
\end{align}
and the 1+3-momentum conservation
\begin{align}
	\tensor{b}{^\mu_\alpha}\Lie{\T u} q_\mu &= - \left(\epsilon + p\right)\tensor[^u]{a}{_\alpha} - \tensor{b}{^\mu_\alpha}\nabla_\mu p - \theta q_\alpha \nonumber \\
	&\quad \, - \tensor{b}{^\mu_\alpha} \nabla_\nu \tensor{\pi}{^\nu_\mu}. \label{eq::1+3-Momentum}
\end{align}

\subsection{Construction of the 4D-Newton equations}
\label{sec::4D_Newton}

In Sec.~\ref{sec::3+1_GR} we detailed the construction of the 3+1-Einstein equations on a time parametrised 3D-manifold $\Sigma$. In the present section we will reverse this construction for the case of the Newton equations: from the parametrised manifold $\SNew$, we will define a spacetime manifold $\MNew$ and write the Newton equations on this manifold.

\subsubsection{Push-forward of the Newton equations}
\label{sec::push_for}

In order to write the Newton equations as equations living in a 4D-manifold $\MNew$, we reverse the pull-back operation of Sec.~\ref{sec::fol_adapt}. The operation $\SNew \rightarrow \MNew$ is called a \textit{push-forward} of $\SNew$ in $\MNew$. While the pull-back in general relativity defined the parametrised manifold $\Sigma$, the push-forward here will define the spacetime manifold $\MNew$.

The push-forward is parametrised by $t$. It hence defines a set $\folNew$ of hypersurfaces embedded in $\MNew$. At this stage, for a general push-forward, these hypersurfaces can intersect. We however impose that the family $\folNew$ defines a foliation in $\MNew$. We note $\T n$ and $\Nfol$ the normal vector and the lapse of this foliation.

In the 3+1-Einstein equations~\eqref{eq::T_Ricci_3D} and~\eqref{eq::T_Ray_3D} on $\Sigma$, the partial time derivative $\derivtn{\T\beta}$ carries the information on the shift of the adapted coordinates in which the pull-back was made. This is not the case for the classical 3D-Newton equations~\eqref{eq::def_theta_omega_New}-\eqref{eq::Wg_New} as no pull-back is at their origin. This means that the derivative $\derivtN{0}$ in the Newton equations does not necessarily correspond to the derivative $\derivtn{0}$ in $\MNew$. Instead in a general push-forward of the Newton equations, $\derivtN{0}$ becomes $\derivtn{\T\B}$, where $\T \B$ is a spatial vector relative to the foliation $\folNew$.

\begin{figure}[h]
	\centering
	\includegraphics[width=\textwidth]{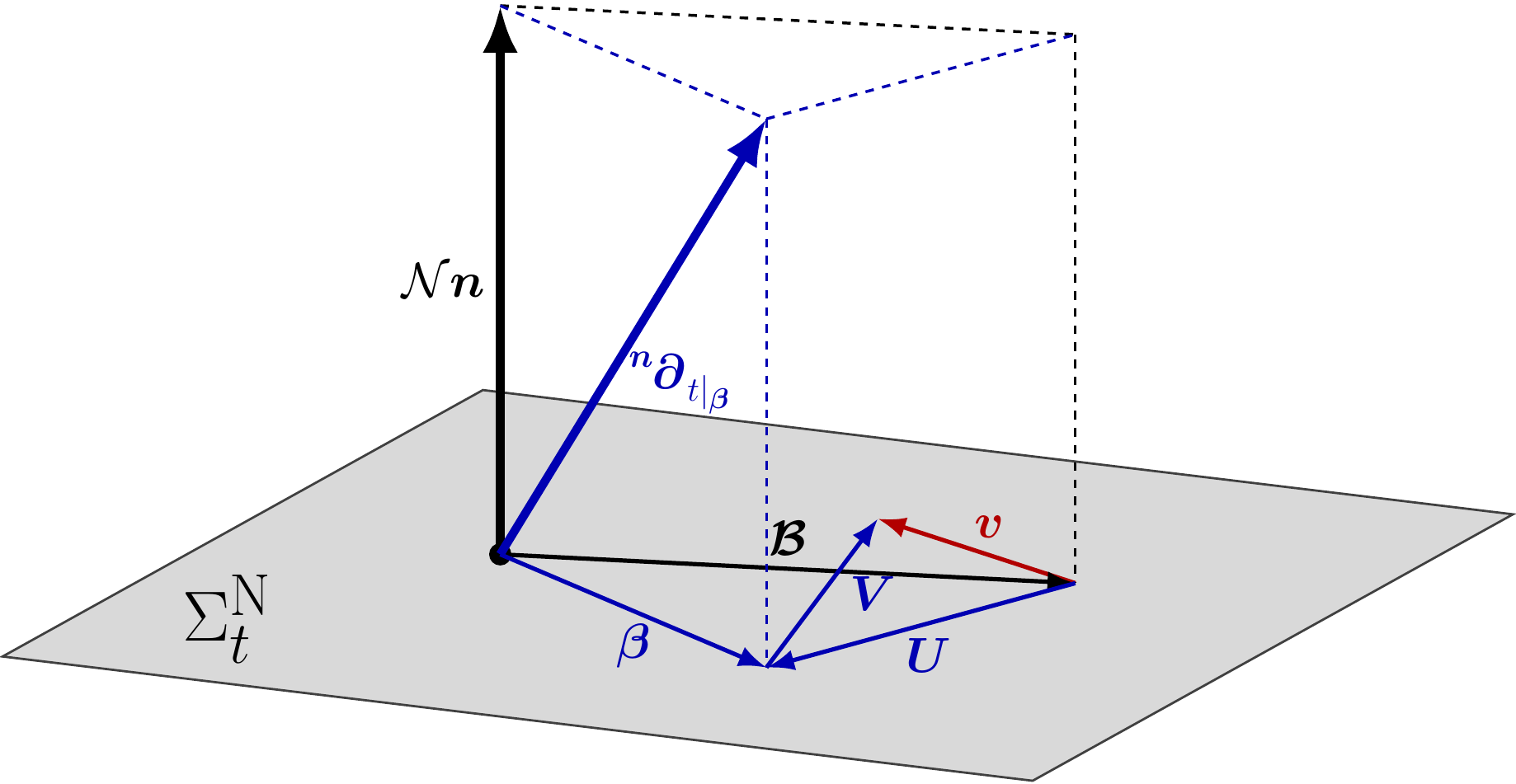}
	\caption{Representation of a slice $\SNew_t$ of the foliation $\folNew$. We show the vectors defining the 4D-manifold $\MNew$ (in black); the Newtonian fluid velocity (in red); the vectors relative to a general adapted class $\classn{\T \beta}$ (in blue).}
	\label{fig::class}
\end{figure}

So a class $\classN{\T \U}$ in $\SNew$ corresponds to an adapted class $\classn{\T \B + \T \U}$ in $\MNew$. This is schematised in Fig.~\ref{fig::class} where we represent a slice $\SNew_t$ and the vectors $\T n$, $\T \B$ and $\T \vN$. We also represent in blue the shift $\T \beta$ and the time vector $\tensor[^{\T \beta}]{\T \partial}{_t}$ of a general adapted class $\classn{\T \beta}$ as well as the vectors $\T \U$ and $\T \V$ defined in Sec.~\ref{sec::Euler-Newton_Pv} with respect to this class.

The only constraint on the foliation $\folNew$, and so on $\MNew$, is to be spatially flat and to have an adapted coordinate system in which the spatial components of the spatial metric do not depend on time. This coordinate system is $\classn{\T \B}$. There are however no constraints on $\Nfol$ or $\T \B$ from the Newton equations.

In the coordinates $\classn{\T \B}$, the spacetime metric is
\begin{align}
	g_{\alpha\beta} = \left(\begin{array}{c|c} \pm \Nfol^2 + \B_c\B^c & \B_b \\ \hline \B_a & h_{ab}(x^c)\end{array}\right), \label{eq::metric_4D}
\end{align}
where $h_{ab}(x^c)$ are the spatial components of the flat spatial metric in the chosen adapted coordinates. The $\pm$ sign depends on the choice of signature for the metric: $+$ for $(++++)$ signature and $-$ for $(-+++)$ signature. This is discussed in Sec.~\ref{sec::sig_New}. As $\MNew$ is determined by the metric~\eqref{eq::metric_4D}, then the choice of $\Nfol$ and $\T \B$ determines this manifold. \\

\remark{The push-forward $\derivtN{0} \rightarrow \Lie{\Tderivtn{\T\B}}$ is only possible if the derivative is applied on a contravariant tensor as $\Lie{\Tderivtn{\T\B}}$ applied on a covariant tensor is not spatial (see appendix~\ref{app::Lie}). This is also true for $\tensor[^\SNew]{{\Lie{\T \vN}}}{} \rightarrow \tensor[^\MNew]{{\Lie{\T \vN}}}{}$, where $\tensor[^\SNew]{{\Lie{\T \vN}}}{}$ and $\tensor[^\MNew]{{\Lie{\T \vN}}}{}$ are respectively the Lie derivative in $\SNew$ and in $\MNew$. Therefore, the push-forward of the Newton-vorticity equation~\eqref{eq::Wg_New} has to be done when written in the contravariant form.}

\subsubsection{4D-Newton equations}
\label{sec::4D-New_eq}

The push-forward on $\MNew$ of the 3D-Newton equations~\eqref{eq::def_theta_omega_New}-\eqref{eq::g_def} gives the 4D-Newton equations
\begin{align}
	\Lie{\Nfol\T n + \T \B + \T \vN} \,&\rho \quad = -\rho \theta, \label{eq::Cont_New4D} \\
	\Lie{\Nfol\T n + \T \B + \T \vN} \, &\theta \quad = - 4\pi G \rho + \Lambda - \Theta_{\mu\nu} \Theta^{\mu\nu} + \omega_{\mu\nu}\omega^{\mu\nu}, \label{eq::Ray_New4D} \\
	\Lie{\Nfol\T n + \T \B + \T \vN} \, &\omega^{\alpha\beta} = - 4\omega^{\mu[\alpha}{\Theta^{\beta]}}_\mu, \label{eq::Wg_New4D}
\end{align}
with the definition of the gravitational field
\begin{align}
	g^\alpha := \Lie{\Nfol\T n + \T \B + \T \vN} v^\alpha + v^\mu\left({\Theta_\mu}^\alpha + {\omega_\mu}^\alpha\right), \label{eq::g_def4D}
\end{align}
where $\Theta_{\alpha\beta}$ and $\omega_{\alpha\beta}$ are defined as
\begin{align}
	\Theta_{\alpha\beta} := \D_{(\alpha}\vN_{\beta)} \quad ; \quad \omega_{\alpha\beta} := \D_{[\alpha}\vN_{\beta]}. \label{eq::kinematical_4DNew}
\end{align}
The constraints on the foliation are that the Ricci tensor of the hypersurfaces $\SNew_t$ is zero for all $t$ and that their extrinsic curvature is
\begin{align}
	K_{\alpha\beta} := \Nfol \D_{(\alpha} {\B}_{\beta)}. 
\end{align}
This amounts to saying that the spatial components of the spatial metric in the coordinates $\classn{\T \B}$ do not depend on time. We call such a foliation, a \textit{Newtonian foliation}.

The system~\eqref{eq::Cont_New4D}-\eqref{eq::kinematical_4DNew} is equivalent to the original system~\eqref{eq::def_theta_omega_New}-\eqref{eq::g_def}, i.e. both systems can be derived from the other. The solutions for $\T\vN$ in the 4D-system are then the same as for the original system. Furthermore, it is still possible to write the 4D Newton-Raychaudhuri equation~\eqref{eq::Ray_New4D} like the Newton-Gauss equation~\eqref{eq::Ray_ENew}. This means that we have the relation $D_\mu g^\mu = -4\pi G\rho + \Lambda$ and this for any choice of $\Nfol$ and $\T \B$. The same applies for the 4D Newton-vorticity equation~\eqref{eq::Wg_New4D} which can be written as $D_{[\alpha}g_{\beta]} = 0$.

As said before, the only constraint at that point on $\MNew$ is to have a Newtonian foliation. So in the general case where $\Nfol$ and $\T \B$ are not chosen, $\MNew$ is not influenced by the dynamics of $\T \vN$. However, choices on $\Nfol$ and $\T \B$ can be made such that the properties of this manifold will depend on $\T \vN$. Such a choice is the subject of Sec.~\ref{sec::Ortho}. Also in Sec.~\ref{sec::Hubble_flow_New_4D} we discuss a choice where $\MNew$ is a homogeneous expanding background manifold. \\

\remark{Making the push-forward from the Newton equations in $\classN{\T\U}$ [Eqs.~\eqref{eq::Cont_NewP}-\eqref{eq::g_defP}] is equivalent as from the same equations in $\classN{0}$, which is done in this section. The equations from $\classN{\T\U}$  are obtained from~\eqref{eq::Cont_New4D}-\eqref{eq::g_def4D} by changing $\T\B$ into $\T\B + \T\U$.}

\subsubsection{Signature of $\MNew$}
\label{sec::sig_New}

While constructing $\MNew$ with its metric given by~\eqref{eq::metric_4D}, we made no assumptions on its signature. The push-forward manoeuvre made in Sec.~\ref{sec::push_for} is independent of this signature. So the metric of $\MNew$ can be either of Lorentzian $(-+++)$, or Euclidean $(++++)$ signature. It is an additional freedom to $\Nfol$ and $\T\B$ of the 4D-Newton equations. We will however take only Lorentzian manifolds. The only argument to take such manifolds is to enable us to directly compare $\MNew$ with solutions of the Einstein equations.

The Lorentzian choice might seem in contradiction with the Galilean invariance of Newton's theory. This is only the case if we ask the connection related to the metric~\eqref{eq::metric_4D}, defined on the spacetime manifold $\MNew$, to have this invariance. This property is however not imposed by the axioms of the classical formulation of Newton's theory on a 3D-manifold, i.e. the one presented in Sec.~\ref{sec::Newton}. In this formulation, no spacetime manifold is defined. That is why, when constructing $\MNew$ from the classical formulation, some freedom appears on the properties of this manifold. This view is different from the Newton-Cartan theory, where the structure on the manifold, defined by two degenerate metrics and a compatible connection, is imposed to be invariant under Galilean transformations. This structure is called a Galilei structure, and the related manifold, a Galilei manifold (see Ref.~\cite{1976_Kunzle}). \\

The Lorentzian choice might also seem in contradiction with the fact that there is no speed limit in Newton's theory, something linked to the notion of causality. We clarify why there is no such contradiction hereafter.

The causality is the relationship between causes and effects of an event, or observer. Thus this notion depends on the definition of observers. In general relativity, the manifold of work is a 4D-manifold, on which an observer is defined by a 4-vector such that the spatial velocity of an event he measures in his rest frames cannot be greater than $c$. This implies that the spacetime manifold is a Lorentzian manifold and that the 4-vector of this observer is a unit, timelike vector.

In the classical formulation of Newton's theory, an observer is described by a velocity vector field $\T p$ in the Euclidean 3-space not limited by the speed of light. If we push-forward this observer in $\MNew$, it is still defined by $\T p$ which is spatial. Then an observer in the 4D-formulation of Newton's theory is not described by a unit 4-velocity vector, but by a spatial vector field, not limited by $c$. On the one hand, contrary to general relativity, this definition of an observer does not require $\MNew$ to be Lorentzian; reversely, choosing $\MNew$ to be Lorentzian does not impose constraints on the definition of an observer in 4D-Newton. On the other hand, this Newtonian definition of an observer, and therefore of causality, naturally allows for the measure of superluminous velocities on the foliation $\folNew$ by any observers as their spatial velocities can themselves be arbitrarily large. \\


In the next subsection we will see that it is possible to physically define a 4-velocity vector $\TuN$ for the Newtonian fluid. We will however necessarily have an additional constraint if we want this vector to be a unit vector [see Eq.~\eqref{eq::cond_vN}]. \\

\remark{The push-forward used to construct $\MNew$ is taken from the 3+1-formalism of GR; it thus automatically defines a spacetime metric, implying $\MNew$ to be (pseudo)-Riemannian. It might however be allowed to use a push-forward which does not necessarily lead to such a manifold. Recovering the Newton-Cartan theory using the method of Sec.~\ref{sec::4D_Newton} should in this case be possible.}

\subsubsection{Newtonian 4-velocity}
\label{sec::TuN}

The equations of Sec.~\ref{sec::4D-New_eq} describe the evolution of a Newtonian fluid in $\MNew$. This fluid is defined by the spatial vector $\T \vN$. We would like to define a vector $\TuN$ which we can call the 4-velocity of the Newtonian fluid. The definition of this vector is not constrained by the 4D-equations, so it remains a choice.

The choice we make is physically motivated by the definition of the Lagrangian coordinates (see Sec.~\ref{sec::Lag_New}). In general relativity, Lagrangian coordinates are defined to be comoving with the fluid 4-velocity, i.e. $\T \partial_t \propto \T u$. For a foliation defined by the normal vector $\T n$ and the lapse $N$, and a tilt velocity vector $\T w$ of the fluid with respect to that foliation, the Lagrangian coordinates correspond to the adapted class $\mathcal{X}^{\T n}_{N \T \vt}$.

In the classical Newton theory, these coordinates correspond to the class $\mathcal{X}_{\T \vN}$ on $\SNew$. Its equivalent on the foliation $\folNew$ is the adapted class $\mathcal{X}^{\T n}_{\T \B + \T \vN}$ (see Sec.~\ref{sec::push_for}). Then we demand that the tilt velocity vector of $\uN$ with respect to the foliation defined by $\T n$ and $\Nfol$ be $\frac{1}{\Nfol}\left(\T \B + \T \vN\right)$.

However there remains a freedom on the choice of the normal part of $\TuN$ with respect to $\folNew$. Two natural choices are possible:
\begin{itemize}
	\item \textit{The Newton-Cartan choice}: this 4D theory features a 1-form $\T\psi$\footnote{Using the notation of K\"unzle~\cite{1976_Kunzle}.} which defines an absolute time and a foliation. An observer in this theory, described by a vector $\T u$, is defined with respect to this absolute time. The vector $\T u$ has then the following property $\psi_\mu u^\mu = 1$. The analogue to this definition in our case would be to impose $n_\mu \uN^\mu = -1$. This leads to a first definition of $\TuN$:
	\begin{align}
		\TuN := \frac{1}{\Nfol}\left(\Nfol \T n + \T \B + \T \vN \right), \label{eq::def_U_N_1}
	\end{align}
	\item \textit{The relativistic choice}: in GR, an observer is described by a unit vector $\T u$, with $u_\mu u^\mu = -1$. In our case, this translates  into $\uN_\mu \uN^\mu = -1$. This leads to a second natural definition of $\TuN$:
	\begin{align}
		\TuN := \frac{\gamma}{\Nfol}\left(\Nfol \T n + \T \B + \T \vN \right), \label{eq::def_U_N_2}
	\end{align}
with $\gamma := \left[1-\frac{1}{\Nfol^2}\left(\B_\mu + \vN_\mu\right)\left(\B^\mu + \vN^\mu\right)\right]^{-1/2}$. The downside of this definition is that it adds the following constraint to the 4D-Newton equations~\eqref{eq::Cont_New4D}-\eqref{eq::kinematical_4DNew}:
	\begin{align}
		\left(\B_\mu + \vN_\mu\right)\left(\B^\mu + \vN^\mu\right) < \Nfol^2. \label{eq::cond_vN}
	\end{align}
	This is indeed a constraint, as if we take $\Nfol = 1$ and $\T\B = 0$, Eq.~\eqref{eq::cond_vN} imposes $\vN_\mu \vN^\mu < 1$. Such a constraint is not implied by the first definition~\eqref{eq::def_U_N_1}.
\end{itemize}

\begin{figure}[h]
	\centering
	\includegraphics[width=\textwidth]{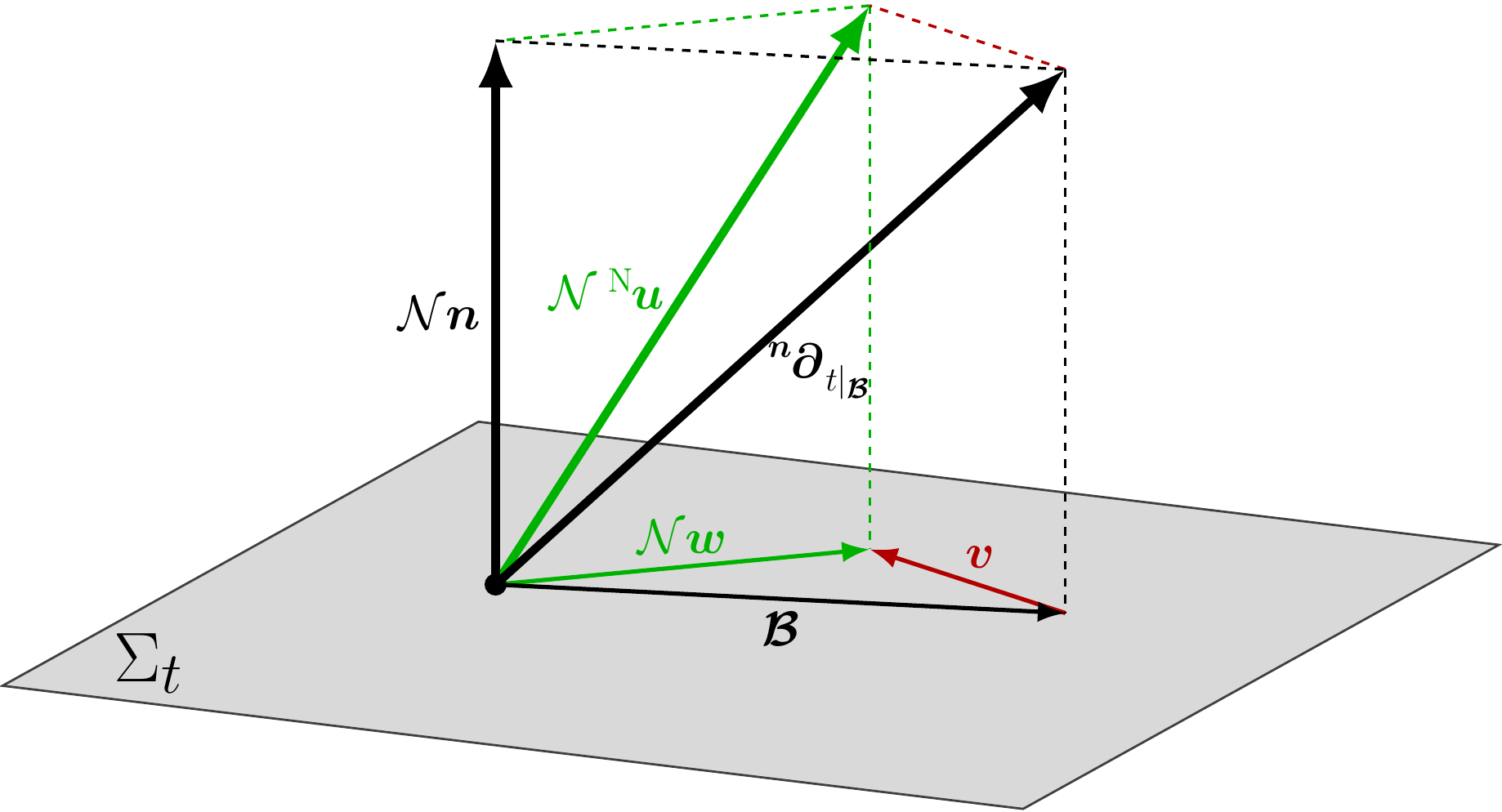}
	\caption{Representation of the chosen Newtonian 4-velocity vector $\TuN$ with respect to the foliation $\folNew$. Using the analogue between Lagrangian coordinates in Newton and in GR, we imposed the tilt velocity of $\TuN$ to be $\T\omega := \frac{1}{N}\left(\T \B + \T \vN\right)$. The normal part is chosen to be $\T n$, which is the equivalent to what is taken in the Newton-Cartan theory for the 4-velocity of an observer.}
	\label{fig::TuN}
\end{figure}

We take the first definition~\eqref{eq::def_U_N_1}, as it remains general with respect to the 4D-Newton equations. This 4-velocity is illustrated in Fig.~\ref{fig::TuN}, along with $\T n$, $\T\B$ and $\T\vN$.

With this choice, we can interpret the 4-velocity $\TuN$ as follows: $\TuN$ corresponds to the covered distance $\Delta x^\mu$ in spacetime per unit of proper time $\tensor[^{\T n}]{\tau}{}$, where $\tensor[^{\T n}]{\tau}{}$ refers to the proper time of $\T n$. This vector $\T n$ and its induced foliation then define a fundamental time (as in the NC theory) with respect to which Newtonian 4-velocities are defined. The situation is different in general relativity, where the 4-velocity of a fluid element is defined as the covered distance $\Delta x^\mu$ in spacetime per unit of fluid element proper time $\tensor[^{\T u}]{\tau}{}$. \\

\remark{With what precedes, we can complete the definition of \textit{a Newtonian observer} in the 4D-Newton theory, as being described by a vector $\T m$ such that $m^\mu n_\mu = -1$. This is the equivalent definition of an observer in the NC theory. The observer given by $\TuN$ is then the fluid itself.}

\remark{For both definitions, $\T \vN$ corresponds to the coordinate velocity of the fluid 4-velocity $\TuN$ with respect to the coordinates $\classn{\T \B}$ (see Fig.~\ref{fig::TuN}). The tilt velocity is however still not the physical vector. The latter remains $\T\vN$ as taking $\T\vN = 0$ still implies a constraint on $\rho$ with the 4D equation~\eqref{eq::Cont_New4D}. This is not the case if we take $\frac{1}{\Nfol}\left(\T \B + \T \vN\right) = 0$.}

\subsubsection{Background homogeneous expanding spacetime}
\label{sec::Hubble_flow_New_4D}

In this section, we present a first choice for the manifold $\MNew$.

Taking $\Nfol := 1$ and $K_{\alpha\beta} = \D_{(\alpha} \B_{\beta)} := - H_{\alpha\beta}$, with $\D_\mu H_{\alpha\beta} := 0$, implies that $\MNew$ is a homogeneous globally expanding spacetime. This expansion is anisotropic, unless $H_{\alpha\beta} \propto h_{\alpha\beta}$ which corresponds to the Einstein-de Sitter spacetime.

The tilt velocity of $\TuN$ is then $\T \vt = \T \B + \T \vN$. The expansion tensor can be rewritten $\Theta_{\alpha\beta} := H_{\alpha\beta} + \D_{(\alpha} \vt_{\beta)}$. Then, in Eulerian comoving coordinates, the 4D-Newton equations for the vector $\T \vt$ become the usual Newton equations with a homogeneous deformation (equations for $\T\V$ introduced in Sec.~\ref{sec::Euler-Newton_PV} with the homogeneous deformation of Sec.~\ref{sec::Hubble_flow_New}). 

We still have the same results concerning expansion in a compact topology. If we impose the hypersurfaces $\SNew_t$ to have a compact topology, then $H_{\alpha\beta}$, being a constant gradient, has to be zero. It is still not possible to have an expanding compact topology in Newton, even when using the 4D-Newton formalism. This was expected as the two formulations are equivalent. In order to do it, the trick is to consider periodic boundaries only on the vector $\T \vt$ as explained in Sec.~\ref{sec::Hubble_flow_New}. In this case the topology of the hypersurfaces $\SNew_t$ is still $\mathbb{R}^3$ as $\T \B$ is not periodically defined.

The choice of 4D-manifold $\MNew$ of this section is independent of the fluid kinematical quantities. It is then only a background manifold. We therefore cannot draw any dictionary between the Newtonian fluid quantities and the relativistic fluid quantities defined via the Einstein equation for the 4D-manifold. In the next section, $\MNew$ will depend on the Newtonian fluid enabling, the definition of a dictionary in Sec.~\ref{sec::dico}. \\

\remark{The choice we make here cannot strictly be called a foliation choice as this would imply that another choice would describe the same equations but in another foliation, the 4D-manifold being unchanged. This is not true as, in general, another choice for $\Nfol$ and $\T \B$ changes $\CM^\mathrm{New}$.}

\section{1+3-Newton equations}
\label{sec::Ortho}

\subsection{The choice}
\label{sec::Ortho_choice}

A natural choice coming from the definition of the Newtonian 4-velocity~\eqref{eq::def_U_N_1} is to take $\Nfol$ and $\T \B$ such that the foliation is orthogonal to $\TuN$, implying $\TuN= \T n$. This is done by taking $\T \B = - \T \vN$. The lapse $\Nfol$ remains unknown. In analogy with GR, as we deal with a dust fluid, we choose the 4-acceleration of $\TuN$ to be zero, which is imposed by $\Nfol = 1$. We expect this choice to be different in the case of non-dust fluids (this is discussed in Sec.~\ref{sec::non-dust}). \\

\remark{Interestingly, with the above choice, the two definitions~\eqref{eq::def_U_N_1} and~\eqref{eq::def_U_N_2} are equivalent.}

Under the present choice the 4D-Newton equations become
\begin{align}
	\Lie{\TuN} \,&\rho \quad = -\rho \theta, \label{eq::Cont_New4D_Orth} \\
	\Lie{\TuN} \, &\theta \quad = - 4\pi G \rho + \Lambda -\Theta_{\mu\nu} \Theta^{\mu\nu} + \omega_{\mu\nu}\omega^{\mu\nu}, \label{eq::Ray_New4D_Orth} \\
	\Lie{\TuN} \, &\omega_{\alpha\beta} = 0, \label{eq::Wg_New4D_Orth}
\end{align}
with the definition of the gravitational field
\begin{align}
	g^\alpha := \Lie{\TuN} v^\alpha + v^\mu\left({\Theta_\mu}^\alpha + {\omega_\mu}^\alpha\right),\label{eq::g_def4D_Orth}
\end{align}
where
\begin{align}
	\Theta_{\alpha\beta} &:  = \D_{(\alpha}\vN_{\beta)} \quad ; \quad \omega_{\alpha\beta} := \D_{[\alpha}\vN_{\beta]}, \label{eq::def_theta_omega} \\
					&\ = \nabla_{(\alpha} \uN_{\beta)},
\end{align}
and with $\uN^\mu \nabla_\mu \uN^\alpha = 0$ and $\nabla_{[\alpha} \uN_{\beta]} = 0$, so that $\TuN$ defines a foliation. Note that the covariant form of the vorticity equation is now possible as only the normal vector remains in the Lie derivative.

The gravitational field definition~\eqref{eq::g_def4D} can be rewritten as
\begin{align}
	g^\alpha := \uN^\mu \nabla_\mu v^\alpha. \label{eq::def_g_acc}
\end{align}
The RHS is spatial as $\TuN$ has no 4-acceleration. We see that the gravitational field corresponds to the 4-acceleration, with respect to the observer $\TuN$, of the Newtonian fluid velocity $\T \vN$. \\

The properties of the foliation $\folNew$ are now linked to those of the fluid with the relation $K_{\alpha\beta} = -\Theta_{\alpha\beta}$. Then the 4D-Newton equations~\eqref{eq::Cont_New4D_Orth}-\eqref{eq::Wg_New4D_Orth} closely resemble the 1+3-Einstein equations~\eqref{eq::Ray_1+3}-\eqref{eq::1+3-Energy} for a dust fluid: on a formal aspect and on the fact that they are expressed in the rest frames of the fluid. We call them \textit{the 1+3-Newton equations}. \\

The main difference between the 1+3-Newton and the 1+3-Einstein equations remains in the definition of the vorticity. In the Einstein equations, it is defined as the antisymmetric rest frame projection of the gradient $\T\nabla {\T u}$ [definition~\eqref{eq::def_omega_GR}]. However in Newton, the antisymmetric part of $\T\nabla\TuN$ is zero as $\TuN$ defines a foliation. Instead the vorticity is defined as the antisymmetric part of a spatial vector gradient [second equation in~\eqref{eq::def_theta_omega}], the symmetric part of that gradient being the expansion tensor [first equation in~\eqref{eq::def_theta_omega}]. This is the reason why we will define the Newtonian limit (see Sec.~\ref{sec::dic_phys}) and the Newton-GR dictionary (see Sec.~\ref{sec::dico}) for irrotational flows. \\

With the choice made in the present section, the spacetime metric of the manifold $\MNew$ in the adapted coordinates $\classn{-\T\vN}$ is
\begin{align}
	g_{\alpha\beta} = \left(\begin{array}{c|c} - 1 + \vN_c\vN^c & \vN_b \\ \hline \vN_a & h_{ab}(x^c)\end{array}\right). \label{eq::metric_1+3-Newton}
\end{align}
where $h_{ab}(x^c)$ are the spatial components of the flat spatial metric in the fixed coordinates used to derive the solution for $\T\vN$. \\

\remark{As said previously, the norm of the Newtonian spatial velocity $\T \vN$ is not bounded by $c$. Where $\T \vN$ is superluminal, the time vector $\Tderivtn{-\T\vN}$ is spacelike and the points in $\MNew$ where $\vN_\mu \vN^\mu = c^2$ correspond to coordinate singularities of the class $\classn{\T \B}$. As we will see in Sec.~\ref{sec::Scwharz}, this is not necessarily unphysical.}

\subsection{3+1-Newton equations}
\label{sec::3+1-Newton}

Once we have chosen $\Nfol$ and $\T \B$, the manifold $\MNew$ is set. The choice made in Sec.~\ref{sec::Ortho_choice}, leading to Eqs.~\eqref{eq::Cont_New4D_Orth}-\eqref{eq::g_def4D_Orth}, is such that these equations are written with respect to the foliation orthogonal to the Newtonian fluid 4-velocity $\TuN$ we defined. This is why they are called 1+3-Newton equations. We can however change this foliation.

We define a timelike unit vector field $\T m$ on $\MNew$, defining a foliation $\{\Sigma^{\mathrm{N},m}_t\}_{t\in \mathbb{R}}$ of lapse $M$ in $\MNew$. We can then decompose $\TuN$, $\T\Theta$ and $\T\omega$ with respect to $\{\Sigma^{\mathrm{N},m}_t\}_{t\in \mathbb{R}}$. The same can be done for the 1+3-Newton equations. Equations~\eqref{eq::Cont_New4D_Orth} and \eqref{eq::Ray_New4D_Orth} are scalar equations and do not need to be projected, contrary to Eqs.~\eqref{eq::Wg_New4D_Orth} and \eqref{eq::g_def4D_Orth}. As for the Lie derivative $\Lie{\TuN}$, it becomes $\Lie{\gamma\T m + \gamma\T w}$ with the usual decomposition of $\TuN$ with respect to $\T m$ defined in~\eqref{eq::tilt_vel}.

Then writing the 1+3-Newton equations in terms of the variables $\TuN$, $\T\Theta$ and $\T\omega$ projected with respect to $\{\Sigma^{\mathrm{New},m}_t\}_{t\in \mathbb{R}}$ gives the 3+1-Newton equations. We do not give these equations here but discuss in Sec.~\ref{sec::Dic_vort_tilt} a possible use of them in relation with dictionary defined in Sec.~\ref{sec::dico}.

\subsection{1+3-Newton from GR}
\label{sec::dic_phys}

The choice of $\Nfol$ and $\T\B$ leading to the 1+3-Newton equations implies that the properties of the Lorentzian manifold $\MNew$ depend on the dynamics of the Newtonian velocity $\T\vN$. We however do not know at which point $\MNew$ with the metric~\eqref{eq::metric_1+3-Newton} is solution of the Einstein equations for the same fluid as the one in 1+3-Newton, i.e. a dust fluid. In this section we will recover the 1+3-Newton equations from GR, enabling us to answer this question in Sec.~\ref{sec::dico}.

\subsubsection{Expansion tensor decomposition}
\label{sec::hyp}

We want to recover the 1+3-Newton equations from general relativity, thus defining a Newtonian limit. Our approach will need the definition of a flow orthogonal foliation. However, as explained in Sec.~\ref{sec::Ortho_choice}, the difference in the definition of vorticity between Newton and GR implies that such a foliation cannot be built in the latter theory as opposed to the former. So we expect that recovering the 1+3-Newton equations from the 1+3-Einstein equations will be more complicated for vortical flows.

We then only take irrotational fluids in both theories. Note that a solution to allow for vorticity, but still dealing with foliations, is to make the limit between the 3+1-Newton equations (presented in~\ref{sec::3+1-Newton}) and the 3+1-Einstein equations. This will not be studied in this paper but it is discussed in Sec.~\ref{sec::Dic_vort_tilt}. \\

We consider the 3+1-Einstein equations~\eqref{eq::T_Ricci}, \eqref{eq::T_Codazzi} and \eqref{eq::T_Ray} in the orthogonal foliation of an irrotational dust fluid of 4-velocity $\T u$ (they are equivalent to the 1+3-Einstein equations for irrotational flows). The Hamilton constraint~\eqref{eq::T_Gauss} is redundant with the other equations and not needed for the discussion. For this section only we use spatial indices and reintroduce the light speed $c$.

In a cosmological setup, we suppose that we can decompose the expansion tensor into scalar, vector and tensor parts as in standard perturbation theory\footnote{In standard perturbation theory this is done for the spatial metric.}:
\begin{align}
	\Theta_{ab} = \chi h_{ab} + \D_{(a} \vN_{b)} + \Xi_{ab} \label{eq::theta_decomp}
\end{align}
with $\tensor{\Xi}{_c^c} = 0$ and $ \D_c\tensor{\Xi}{^c_a} = 0$. $h_{ab}$ is the spatial metric of the orthogonal foliation. The irreducibility of this decomposition is discussed in appendix~\ref{app::decomp}.

We take $\T\vN$ to be irrotational, i.e. $\D_{[a} \vN_{b]} = 0$. This is a choice motivated by the 1+3-Newton equations in which $\D_{[a} \vN_{b]}$ plays the role of the vorticity. In Sec.~\ref{sec::Dic_vort_orth} we discuss what $\D_{[a} \vN_{b]}$ should be in the case of rotational fluids. \\

\remark{The scalar-vector-tensor decomposition we made is fully covariant (it does not depend on an adapted class of coordinates). It is also independent of a choice of foliation as the spatial projection used is defined with respect to the fluid. This is not the case for the decomposition of the spatial metric in standard perturbation theory.}

The parameter $\chi$ is interpreted as the \textit{scalar expansion},  $\D_{(a} \vN_{b)}$ as the \textit{gradient expansion} and $\T\Xi$ as the \textit{gravitational wave term}. $\T\Xi$ is only a shear term as its trace is zero. While the trace of the gradient expansion is on average zero for a compact space, the scalar expansion is not. Then global expansion in a compact space is driven by $\chi$. Both $\chi$ and $\T\Xi$ are not present in Newton, where only $\D_{(a} \vN_{b)}$ is. This is coherent with the fact that there are no gravitational waves nor global expansion in a compact space for this theory. In this view, we can then interpret $\T\D\T\vN$ to be the \textit{Newtonian fluid expansion} and $\chi$ to be the \textit{space expansion}.

\subsubsection{The limit}
\label{sec::the_lim}

The first approximation we make is to neglect the space expansion and the gravitational wave term compared to the Newtonian fluid expansion (this is discussed in appendix~\ref{app::approx_D_D}):
\begin{align}
	\Theta_{ab} \simeq \D_{a}\vN_{b}. \label{eq::hyp}
\end{align}
This implies that in the adapted class $\class{-\T\vN}{\T u}$
\begin{align*}
	\derivt{-\T\vN}{\T u} h_{ab} \ll \D_{a} \vN_{b},
\end{align*}
and thus
\begin{align}
	\derivt{-\T\vN}{\T u} \left( \D_{a}\vN_{b} \right) \simeq  \D_{a} \left(\derivt{-\T\vN}{\T u} \vN_{b}\right). \label{eq::dic_commut}
\end{align}
In this commutation relation we neglected the time variation of the spatial metric.

We define the beta-factor $\beta_\vN := |\T \vN|/c$ and the following length scales:
\begin{itemize}
	\item the typical length scale $L_{\vN,l}$ of the spatial variation of the vector $\T \vN$, i.e. $\frac{1}{c}\D_a\vN_b = \frac{1}{c}\Theta_{ab} \sim \beta_\vN/L_{\vN,l}$,
	\item the typical length scale $L_{\vN,t}$ of the time variation of the vector $\T \vN$, i.e. $\frac{1}{c^2}\derivt{-\T\vN}{\T u} v^a \sim \beta_\vN/L_{\vN,t}$,
	\item the Schwarzschild density length scale $L_\epsilon := \left(\frac{G \epsilon}{c^4}\right)^{(-1/2)}$,
	\item the  typical local curvature radius $L_{R}$ of the spatial Ricci tensor, i.e. $R_{ab} \sim 1/L_R^2$.
\end{itemize}
By defining the Newtonian gravitational field as $g^a := (\derivt{-\T\vN}{\T u} + v^c\D_c)v^a$ [justified by the 1+3-Newton equation~\eqref{eq::def_g_acc}], we can say that in a Newtonian regime, $\derivt{-\T\vN}{\T u}  v^a$ will be of the same order as $v^c\D_cv^a$ which implies
\begin{align*}
	L_{\vN,l}/L_{\vN,t} \sim \beta_\vN.
\end{align*}

Assuming that $\frac{1}{c} D_aD_b v_c \sim \beta_\vN/L_{\vN,l}^2$ and using the 3+1-Raychaudhuri equation~\eqref{eq::T_Ray_3D} with the commutation relation~\eqref{eq::dic_commut} we have
\begin{align*}
	\left(\frac{L_{\vN,l}}{L_\epsilon}\right)^2 \sim \beta_\vN^2.
\end{align*}
This relation along with the 3+1-Ricci equation~\eqref{eq::T_Ricci_3D} leads to
\begin{align}
	\left(\frac{L_{\vN,l}}{L_R}\right)^2 \sim \beta_\vN^2. \label{eq::L_v/L_R}
\end{align}
For the usual condition $\beta_\vN \ll 1$ on the Newtonian velocity $\T \vN$, relation~\eqref{eq::L_v/L_R} shows that the spatial variations of the Newtonian velocity are small in front of the typical length scale given by the spatial curvature. This shows that the curvature is of second order in $\beta_\vN$.

We can then consider that, at leading order, the spatial curvature does not affect the dynamics of $\T \vN$, i.e. $\T \D \T \vN \sim \TDh \T \vN$, where $\TDh$ is the connection of a flat metric $\T\hh$ (a more quantitative justification of this approximation is given in appendix~\ref{app::approx_D_D}). Then, at leading order, the 3+1-Ricci equation is not an evolution equation anymore but becomes a relation giving the spatial curvature orthogonal to the fluid as function of the kinematical quantities of that fluid. In this view, we then have $R_{ab} = R^{(2)}_{ab}$, where $R^{(2)}_{ab}$ is of second order in $\beta_\vN$ with
\begin{align}
	R^{(2)}_{ab} = &\frac{-1}{c^2}\Big[\Big(\derivt{-\T\vN}{\T u}  + \tensor[^\Sigma]{{\Lie{\T \vN}}}{}\Big) \Theta_{ab} \label{eq::R_ab^2} \\
	&  + \left(\frac{4\pi G\epsilon}{c^2} + \Lambda\right) \, \hh_{ab} + \theta \Theta_{ab} - 2\Theta_{ac}{\Theta^c}_b\Big], \nonumber
\end{align}
with $\T \Theta = \hat{\T\D}\T \vN$.

As for the momentum constraint~\eqref{eq::T_Codazzi_3D}, it becomes $\hat\D_{[a}\hat\D_{b]} v_c = 0$ at leading order in $\beta_\vN$. This is consistent with a zero curvature at first order. Thus Eq.~\eqref{eq::T_Codazzi_3D} is not a constraint anymore.

We give an additional relation for the Weyl tensor in this limit (see Ref.~\cite{1998_Maartens_et_al} for the expression of the Weyl tensor in terms of the kinematical quantities $\T\Theta$ and $\T\omega$). Its electric part $E_{ab}$ is
\begin{align}
	E_{ab} = -\D_{\langle a}g_{b\rangle}.
\end{align}
This relation is true to any order in $\beta_\vN$ once assumption~\eqref{eq::hyp} is made. The magnetic part $H_{ab}$ is zero. Note that if $\D_{[a} \vN_{b]} \not= 0$, this is not true anymore. This is discussed in Sec.~\ref{sec::Dic_vort_orth}. \\

\remark{Using the decomposition of the expansion tensor~\eqref{eq::theta_decomp}, we can see that the solutions of the Einstein equations which do not feature the Newtonian fluid expansion term will not have a Newtonian limit. In particular, this is the case for purely gravitational waves solutions.}

\subsubsection{Recovering the 1+3-Newton equations}

The limit introduced in the previous section implies that at leading order in $\beta_\vN$ the Ricci equation is a relation for the spatial curvature and not an evolution equation anymore. This spatial curvature orthogonal to the fluid 4-velocity is of second order in $\beta_\vN$. The expansion tensor is a gradient, $\Theta_{ab} = \D_{(a}\vN_{b)}$. The momentum constraint is then trivial at leading order. The only 3+1-Einstein equations remaining to determine the evolution of $\Theta_{ab}$ are the 3+1-Raychaudhuri equation~\eqref{eq::T_Ray} and the 3+1-energy conservation~\eqref{eq::3+1-Energy}, which are respectively equivalent to Eqs.~\eqref{eq::Ray_New4D_Orth} and~\eqref{eq::Cont_New4D_Orth} of the 1+3-Newton system. The Newton-vorticity equation~\eqref{eq::Wg_New4D_Orth} is trivially recovered as the limit is done for irrotational flows.

With the Newtonian limit defined in the previous subsection, we recovered the 1+3-Newton equations in the irrotational case. This formulation of Newton's equations is then supported. In the next section we will use the Newtonian limit of the present section to define a Newton-GR dictionary.

\subsection{Newton-GR dictionary}
\label{sec::dico}

In the previous section we showed that we can recover the 1+3-Newton equations for irrotational fluids from GR with a limit at leading order in $\beta_{\vN}$. The limit also defines a spacetime manifold $\cal M_\mathrm{lim}$ solution of the Einstein equations at leading order. This manifold however differs from the manifold $\MNew$ given by the metric~\eqref{eq::metric_1+3-Newton}. Indeed, $\MNew$ has strictly flat spatial sections orthogonal to the fluid, whereas the curvature of the same sections in $\cal M_\mathrm{lim}$ is non-zero and of second order. This implies that $\MNew$ with the metric~\eqref{eq::metric_1+3-Newton} is not solution of the Einstein equations at leading order. The dictionary will therefore be given by $\cal M_\mathrm{lim}$ and not $\MNew$. For the remainder of this paper $\cal M_\mathrm{lim}$ will then be denoted ${\cal M}^\mathrm{(dic)}$.

We define the following \textit{Newton-GR dictionary for irrotational dust fluids}: given a solution of the Newton equations for $\T\vN$, the relativistic quantities, denoted with the upper-script~$^{(\mathrm{dic})}$, are determined by the following relations:
\begin{align}
	\epsilon^{(\mathrm{dic})}				&= \rho, \label{eq::dic_eps} \\
	\Theta^{(\mathrm{dic})}_{ab}			&= \Dh_a \vN_b, \label{eq::dic_theta} \\
	R^{(\mathrm{dic})}_{ab}				&= -\Dh_a g_b - \Dh_c\vN^c \Dh_a \vN_b + \Dh_a \vN_c \Dh^c \vN_b \nonumber \\
	& \quad \, + \left(4\pi G \rho + \Lambda\right) \hh_{ab}, \label{eq::dic_phys_R_ij}
\end{align}
where the RHS are the Newtonian quantities. $\T\Theta^{(\mathrm{dic})}$ is the expansion tensor of the relativistic fluid, $\T R^{(\mathrm{dic})}$ is the spatial curvature orthogonal to the relativistic fluid, $\T\Dh$ is a flat connection, $\T g$ is the Newtonian gravitational field constrained by the Newton-Gauss equation~\eqref{eq::Ray_ENew} and the Newton-Faraday equation~\eqref{eq::Wg_ENew}. \\

\remark{As we only give the Ricci tensor $R^{(\mathrm{dic})}_{ab}$, the spatial metric orthogonal to the fluid 4-velocity cannot be explicitly constructed. However, raising or lowering the dictionary quantities at leading order only requires the flat spatial metric $\hh_{ab}$.}

Studying the light ray trajectories with a Newtonian solution using our dictionary requires the 3+1 light-geodesic equation of the manifold ${\cal M}^\mathrm{(dic)}$. This equation can be found in Ref.~\cite{2012_Vincent_et_al}. \\

In Sec.~\ref{sec::Scwharz} we will take the example of an exact solution of the Newton equations to test the dictionary.

As the 1+3-Newton equations were recovered from GR only for irrotational fluids, we were only able to draw a dictionary for these kinds of fluids. In Sec.~\ref{sec::Dic_vort} we discuss the possibility of a dictionary with vorticity.

\subsection{Schwarzschild geometry}
\label{sec::Scwharz}

\subsubsection{Point mass Newtonian solution}

In this section we study an exact vorticity-free solution of the 1+3-Newton equations.

We begin with a 3D-Newtonian calculation. We consider a point mass of mass $M$ creating a gravitational field $g^a = (-GM/r^2, 0, 0)$ in spherical coordinates $(r,\theta,\varphi)$. We then solve the Euler equation~\eqref{eq::Euler_ENew} for a \textit{stationary}, irrotational fluid of test observers of velocity $\vN$. We have
\begin{align*}
	\D_a\left(\vN_c \vN^c\right) = 2 g_a, \\
	v_a = \D_a \Psi,
\end{align*}
with $\Psi$ a scalar field depending only on the radial coordinates. The general solution is
\begin{align}
	v^a = \left(\pm\sqrt{2E + \frac{2GM}{r}} , 0 , 0\right), \label{eq::v_free_fall}
\end{align}
where $E$ is a constant corresponding to the energy of the fluid particles. This solution corresponds to a radially ingoing or outgoing free-falling fluid of test observers. If $E<0$, the solution is valid in the region $r < -\frac{2GM}{E}$ and corresponds to fluid particles with bounded orbits, i.e. elliptic orbits. If $E = 0$, the orbits are parabolic and for $E > 0$ they are hyperbolic. Note that all the particles have the same type of orbit as $E$ is a constant of space. \\

\remark{In the case of the 1+3-Newton equations, this solution implies the following spacetime line element for $\MNew$ in the adapted class $\class{-\T \vN}{\TuN}$ (we recall that $\class{-\T \vN}{\TuN}$ corresponds to the class $\classN{0}$):
\begin{align}
	\dd s^2 =	&\left(-1 + 2E + \frac{2G M}{r}\right) \dd t^2  \mp 2 \sqrt{2E + \frac{2GM}{r}} \dd t \dd r \nonumber \\
			&+ \dd r^2  + r^2 \dd \Omega^2, \label{eq::metric_GP_New}
\end{align}
with $\dd \Omega^2 := \dd \theta^2 + \sin^2\theta \dd \varphi^2$.}

\subsubsection{Relativistic quantities from the Newtonian solution}

Using the dictionary~\eqref{eq::dic_eps}-\eqref{eq::dic_phys_R_ij}, we can derive the relativistic quantities corresponding to the Newtonian solution~\eqref{eq::v_free_fall}. We obtain
\begin{align}
	\epsilon^{(\mathrm{dic})}		&= 0, \label{eq::Sc_eps} \\
	\Theta^{(\mathrm{dic})}_{ab}	&= \mathrm{diag}\left( -\frac{GM}{r^2\vN^r}, r \, \vN^r, r \, \vN^r \sin^2\theta\right), \label{eq::Sc_Theta} \\
	R^{(\mathrm{dic})}_{ab}		&= -2E \ \mathrm{diag}\left( 0, 1, \sin^2\theta\right), \label{eq::Sc_Rab}
\end{align}
with $\vN^r := \pm\sqrt{2E + \frac{2GM}{r}}$.

\subsubsection{Radially free-falling test fluids in GR}

We want to know if the manifold defined by Eqs.~\eqref{eq::Sc_eps}-\eqref{eq::Sc_Rab} is solution of the Einstein equations at leading order in $\beta_\vN$ and if it describes the same physical system as the Newtonian solution, i.e. a radially free-falling test fluid on a mass point.

The solution for this physical system in GR is given by the Schwarzschild manifold and the adapted coordinates corresponding to a free-falling observer are the generalised Gullstrand-Painlev\'e coordinates (see MacLaurin~\cite{2019_MacLaurin}\footnote{MacLaurin~\cite{2019_MacLaurin} uses the Killing energy $e := -n_\mathrm{ff}^\mu \, \xi_\mu$ where $\T n_\mathrm{ff}$ is the 4-velocity of the free falling observer and $\T \xi$ is a static Killing vector. The energy definition we use is linked to $e$ with the relation $E = e^2 - 1$.}). The Schwarzschild line element in these coordinates is
\begin{align}
	\dd s^2 =	&\frac{-1 + \frac{2G M}{r}}{2E+1} \dd t^2  \mp 2\frac{\sqrt{2E + \frac{2GM}{r}}}{2E + 1} \dd t \dd r \nonumber \\
			&+ \frac{1}{2E+1}\dd r^2  + r^2 \dd \Omega^2. \label{eq::metric_GP_gen}
\end{align}
where $E$ can be interpreted as the Newtonian energy of the fluid particles. For $E = 0$, the observer associated with the generalised Gullstrand-Painlev\'e coordinates is a parabolic radially free-falling test fluid. For $E<0$ and $E>0$ the free-fall is respectively elliptic and hyperbolic.

We note with the upper-script~$^{(\mathrm{GP})}$ the relativistic quantities corresponding to an observer associated with the generalised Gullstrand-Painlev\'e coordinates. These quantities are
\begin{align}
	\epsilon^{(\mathrm{GP})}		&= 0, \label{eq::Sc_eps_GR} \\
	\Theta^{(\mathrm{GP})}_{ab}	&= \mathrm{diag}\left( -\frac{GM}{r^2\vN^r\left(2E+1\right)}, r \, \vN^r, r \, \vN^r \sin^2\theta\right), \label{eq::Sc_Theta_GR} \\
	R^{(\mathrm{GP})}_{ab}		&= -2E \ \mathrm{diag}\left( 0, 1, \sin^2\theta\right), \label{eq::Sc_Rab_GR}
\end{align}
with $\vN^r := \pm\sqrt{2E + \frac{2GM}{r}}$.

\subsubsection{Comparison}

In the present section, we compare  the relativistic quantities~\eqref{eq::Sc_eps}-\eqref{eq::Sc_Rab} obtained from the Newton-GR dictionary with the ones of the Schwarzschild metric~\eqref{eq::Sc_eps_GR}-\eqref{eq::Sc_Rab_GR}. For simplicity, we will not consider the case $E<0$. Then $E$ corresponds to the Newtonian energy of the test particles at infinity.

The energy densities $\epsilon^{(\mathrm{dic})}$ and $\epsilon^{(\mathrm{GP})}$ are the same. This was expected as the Newtonian and the GR solutions are both vacuum solutions. The covariant components of the spatial curvatures are also the same, with $R^{(\mathrm{dic})}_{ab} = R^{(\mathrm{GP})}_{ab}$. The covariant components of the expansion tensors differ only for the component $_{rr}$, with $\Theta^{(\mathrm{dic})}_{rr} = (2E+1)\Theta^{(\mathrm{GP})}_{rr}$.

The limit under which the dictionary is defined implies $|\T\vN| \ll c$ for all $r$. Taking $r \rightarrow \infty$, this implies $E \ll 1$ which in turn implies that the comparison should be done in the region $r \gg GM$. Then at leading order in $E$ and $\frac{GM}{r}$, the dictionary quantities are the same as those of Gullstrand-Painlev\'e. This supports our dictionary.

\subsubsection{The parabolic free-fall:  $E = 0$}

In the case $E=0$, the dictionary quantities are exactly equal to the general Gullstrand-Painlev\'e ones. Furthermore the metric~\eqref{eq::metric_GP_New} of the manifold $\MNew$, constructed from the Newtonian solution~\eqref{eq::v_free_fall}, is exactly the Schwarzschild metric in generalised Gullstrand-Painlev\'e coordinates, which implies that $\MNew$ is the Schwarzschild manifold. This result is true without any approximation. Then the 4D construction of Newton's equations we introduced in this paper, and in particular the case of the 1+3-Newton equations, allows us to recover exactly a physical solution of the Einstein equations. This further supports the choice $\TuN = \T n$ made in Sec.~\ref{sec::Ortho_choice}.

Note that it was already known that the velocity as a function of the point mass distance of a parabolic radially free-falling observer was the same in Newton and in GR. What we showed is that this solution allows us to recover from Newton the full spacetime metric of Schwarzschild. This was possible because the foliation of the generalised Gullstrand-Painlev\'e coordinates with $E=0$ has flat spatial sections, which is required by the 1+3-Newton equations. \\

We see from this solution that even if the Newtonian velocity $\T \vN$ can, at certain points of $\MNew$, be comparable to the speed of light, and even exceed it, it is still physical. We know that because it is the Schwarzschild spacetime. This means that solutions of the Newton equations are not necessarily unphysical for $\vN^\mu \vN_\mu \sim c^2$. We however expect this statement to be true in few cases only. \\

\remark{Strangely, this exact correspondence between a Newtonian solution and a GR one arises for a Newtonian fluid whose energy is zero for any fluid particles. This leads to the following question: is there a link, in general, between the energy of a Newtonian fluid and the validity of the related Newtonian solution with respect to GR? If this is the case, this would be true only for one gravitational potential energy convention. For an isolated system like a mass point, we saw that it works if the gravitational potential is taken to be zero at infinity. What convention should be taken in the case of a compact spacelike $\mathbb{T}^3$ remains to be determined.}

\section{Discussions}
\label{sec::Disc}

\subsection{Remarks on the Newtonian limit}
\label{sec::New_limit}

The Newtonian limit of general relativity and the corresponding dictionaries (e.g. \cite{2012_Green_et_al}) are usually done with respect to an accelerated observer. We note its 4-velocity $\T n$. The foliation corresponding to this observer has then a lapse $N$, the spatial gradient of which is the acceleration of the observer. Hence in the adapted coordinates $\classn{0}$, the component $g_{00}$ of the spacetime metric is $N^2$. The acceleration of the observer is considered in these limits to be the gravitational field of Newton theory. Using $\tensor[^n]{a}{_\alpha} := \D_\alpha \ln N$, this is why, at leading order, the lapse, and therefore the component $g_{00}$, gives the Newtonian gravitational potential. In this leading order approximation, the accelerated observer is considered to be only slightly tilted with respect to the fluid. \\

The Newtonian limit we defined in Sec.~\ref{sec::dic_phys} is however made in the rest frames of the fluid, which is not accelerated, being a dust fluid. The Newtonian gravitational field then cannot be the lapse, which is fixed to 1. In place, this field is defined as the acceleration of the spatial vector present in the decomposition of the expansion tensor [see Eq.~\eqref{eq::def_g_acc}].

However, the interpretation of $g_{00}$ as the gravitational potential, in coordinates adapted to the fluid rest frames, still holds in the case of stationary irrotational fluids. In these cases,
\begin{align*}
	g_\alpha	&= D_\alpha\left(\frac{1}{2}\vN_\mu\vN^\mu\right).
\end{align*}
In the coordinates $\class{-\T\vN}{\TuN}$, we have $g_{00} = -1 + \vN_\mu\vN^\mu$ which implies 
\begin{align}
	g_\alpha	&= \frac{1}{2}D_\alpha\left(g_{00} + 1\right), \label{eq::g_station}
\end{align}
and $g_{00}$ can be interpreted as the gravitational potential.

Note that this interpretation is not valid in the case of the Lagrangian coordinates $\class{0}{\TuN}$ where $g_{00} = 1$ as all the dynamics of the fluid is put in the time variations of the spatial metric. Furthermore, in the case of non-stationary fluids, the term $g_{00}$ is not the gravitational potential anymore, as~\eqref{eq::g_station} does not hold.

\subsection{Dictionary with vorticity}
\label{sec::Dic_vort}

In Sec.~\ref{sec::dico} we drew our Newton-GR dictionary in the case of irrotational fluids. The reason for this was that, in general relativity, no orthogonal foliation can be defined for a rotational fluid. But as we made the dictionary in the rest frames of the fluid, we needed such a foliation.

We will not detail the construction of a dictionary with vorticity in this article. We however present two possibilities that should allow for it.

\subsubsection{Tilted dictionary}
\label{sec::Dic_vort_tilt}

The first, and most promising possibility, is to make the dictionary in a tilted foliation with respect to the fluid. In general relativity, the 3+1-Einstein equations provide the tilted description of a fluid and allow for vortical flows. In our formulation of Newton's theory, the equations where the fluid is tilted are the 3+1-Newton equations presented in Sec.~\ref{sec::3+1-Newton}. They are derived from the 1+3-Newton equations by making a change of foliation $\TuN \rightarrow \T m$.

One strength of a tilted dictionary would be to show that Newton's theory can be obtained from any foliation\footnote{\label{foot:: bite} In standard perturbation theory, this means that any gauge choice would be suited for a Newton-GR dictionary.}. But as the choice of this foliation is not necessarily physically motivated (see also Ref.~\cite{2020_BMR} for a discussion of this topic), we would prefer making the dictionary with respect to the fluid rest frames. We discuss this in the next subsection.

\subsubsection{Orthogonal dictionary}
\label{sec::Dic_vort_orth}

Constructing a dictionary in the rest frames of the fluid might be more complicated as no foliation can be defined in general relativity, contrary to the Newtonian case.

It is however possible to define a rest frame Riemann tensor $\tensor[^{\T u}]{\T {\mathrm{Riem}}}{}$ and a rest frame covariant derivative $\tensor[^{\T u}]{\T \D}{}$ (see Ref.~\cite{2014_Roy}). They do not have the same properties as the ones defined on hypersurfaces. The first Bianchi identity for $\tensor[^{\T u}]{\T {\mathrm{Riem}}}{}$ will feature the vorticity of the fluid and $\tensor[^{\T u}]{\T \D}{}$ will have torsion. The latter is however of second order in $\beta_\vN$. We then hope that at leading order the rest frames can be approximated to be a family of hypersurfaces.

It remains to be shown that the projection along $\T u$ of the Lie brackets of rest frame vectors is also of second order\footnote{As there is no foliation, there is no coordinate basis in the rest frames of the fluid. This implies that the Lie brackets of any vectors in these rest frames feature a non-zero part along $\T u$.}. This would indicate that we could maybe define a coordinate basis on the rest frames at leading order. \\

\remark{In this dictionary, the gradient in the decomposition~\eqref{eq::theta_decomp} of the expansion tensor would feature a non-zero antisymmetric part which would be the vorticity tensor. Subsequently, the magnetic part of the Weyl tensor would not be zero anymore.}

\subsubsection{Is it really possible?}

It is known that Newton's theory features gravitational phenomena which are not described by GR. Assuming that the latter is the genuine theory of gravitation, these phenomena are not physical. We mentioned in the introduction the case of a shear-free dust fluid which can both rotate and expand in Newton, but cannot in GR. This implies that no limit exists which allows us to recover the full Newtonian theory from GR.

To our knowledge, there exists no example of a phenomenon like the one just mentioned, i.e. present in Newton but not in GR, for a vorticity-free fluid. If this is indeed the case, this might imply that the impossibility at fully recovering Newton from GR, is due to the vorticity. Then constructing a dictionary with vorticity, as we presented in the previous subsections, would need require additional approximation than just $|\T\vN| \ll c$.

\subsection{1+3-Newton for non-dust fluids}
\label{sec::non-dust}

We assumed until now the Newtonian fluid to be a dust fluid. This was done to simplify the interpretations made while constructing the 1+3-Newton equations and the related dictionary. We briefly study the case of a non-dust fluid in this section.

Such a fluid is influenced by additional forces, other than the gravitational force, described by a vector field $\T F$. These forces can be either internal, linked to the fluid properties (density, pressure, viscosity, ...), or external. The changes in the Newton system for a non-dust fluid is given by the second law of Newton. This is translated by the addition of $\T F$ in the Euler equation~\eqref{eq::Euler_ENew}:
\begin{align}
	\left(\partial_{t|_{_{x}}}  + \vN^kD_k\right)  \vN^i = g^i + F^i/\rho,\label{eq::Euler_ENew_non_dust}
\end{align}
with $\T g$ still solution of the Newton-Gauss~\eqref{eq::Ray_ENew} and Newton-Faraday~\eqref{eq::Wg_ENew} constraints.

In the Newton system~\eqref{eq::def_theta_omega_New}-\eqref{eq::Wg_New} written in terms of kinematical quantities of the fluid, the change is made by adding the divergence of $\T F$ in the Newton-Raychaudhuri equation~\eqref{eq::Ray_New} and the vorticity of $\T F$ in the Newton-vorticity equation~\eqref{eq::Wg_New}. These additional terms are then present in the 4D-Newton equations.

The 1+3-Newton equations should not however be obtained with the choice $\Nfol = 1$ and $\T \B = -\T \vN$ but rather with
\begin{align}
	\T\B = - \T\vN \quad ; \quad \T\D \ln \Nfol = \T F.
\end{align}
This choice would not change the Newton-GR dictionary much. Essentially the interpretation of the Ricci equation to be a relation for the spatial curvature tensor would remain valid.

\subsection{Cosmological models from 1+3-Newton}
\label{sec::Post_New}

In the introduction of this paper we motivated the construction of the 1+3-Newton system as a way to better understand why Newton's theory, compared to GR, is lacking the phenomenon of backreaction (in a compact space). Ultimately this would be used to define simple models suited for the study of backreaction and global topology in cosmology. This section aims at presenting how we could define such models from the 1+3-Newton formulation and GR. However, we leave the precise construction of these models for another paper.

In Sec.~\ref{sec::Post_New_exp}, we present an extension of our dictionary to allow for global expansion of a compact space, but still without backreaction. The next two subsections focus on possible strategies enabling the construction of the cosmological models.

\subsubsection{1+3-Newton equations and dictionary for a globally expanding compact space}
\label{sec::Post_New_exp}

In Secs.~\ref{sec::Hubble_flow_New} and~\ref{sec::Hubble_flow_New_4D} we showed that no global expansion is possible in a compact space in Newton's theory. A solution to allow for expansion was to decompose the fluid velocity $\T\vN$ into a homogeneous deformation vector $\T \U$ and a peculiar velocity $\T\V$, the latter having periodic boundary conditions. As explained in Sec.~\ref{sec::Hubble_flow_New}, this is an effective picture of the expansion in a compact space, as $\SNew$ (or equivalently the hypersurfaces $\SNew_t$ for the 4D formulation) is still $\mathbb{R}^3$.

Having $\SNew$ compact with a global expansion is possible with a modification of the Newton equations based on the effective picture of Sec.~\ref{sec::Hubble_flow_New} and the decomposition introduced in Sec.~\ref{sec::hyp}. We will focus on a modification allowing for isotropic global expansion. \\

The modification is to replace the definition~\eqref{eq::def_theta_omega} for the expansion tensor by
\begin{align}
	\Theta_{\alpha\beta} := H h_{\alpha\beta} + \D_{(\alpha} \vN_{\beta)}. \label{def::theta_dic_H}
\end{align}
where $H$ is a homogeneous Hubble expansion rate (i.e. $\D_\alpha H = 0$), while still using the 1+3-Newton equations~\eqref{eq::Cont_New4D_Orth}-\eqref{eq::g_def4D_Orth}. $H$ an additional fundamental variable in the theory. These equations, along with the definition~\eqref{def::theta_dic_H}, are equivalent to the Hubble flow equations of Sec.~\ref{sec::Hubble_flow_New} but allow $\SNew_t$ to be compact. Note that the evolution equation for $H$, being a spatial constant, is given by the spatial average of the Raychaudhuri equation over the whole manifold $\SNew$\footnote{In the case $\SNew = \mathbb{R}^3$, the spatial average requires boundary conditions at infinity to be defined.}. This average equation then depends on the boundary conditions at infinity if $\SNew = \mathbb{R}^3$ or on the topology if $\SNew$ is compact. \\

We can then redefine the dictionary of Sec.~\ref{sec::dico} to feature the global expansion. We then have a \textit{Newton-GR dictionary for irrotational dust fluids and globally expanding compact spaces}:
\begin{align}
	\epsilon^{(\mathrm{dic})}				&= \rho, \label{eq::dic_eps_H} \\
	\Theta^{(\mathrm{dic})}_{ab}			&= \Theta_{ab}, \label{eq::dic_theta_H} \\
	R^{(\mathrm{dic})}_{ab}				&= -\Dh_a g_b - \theta \Theta_{ab} + \Theta_{ac}\tensor{\Theta}{^c_b} + \left(4\pi G \rho + \Lambda\right) \hh_{ab}, \label{eq::dic_phys_R_ij_H}
\end{align}
The RHS side are the Newtonian quantities with $\Theta_{ab} := H h_{ab} + \Dh_{a} \vN_{b}$ and $\Dh_{[a} \vN_{b]} = 0$. \\

The modification~\eqref{def::theta_dic_H} can be justified by the limit introduced in Sec.~\ref{sec::the_lim} and appendix~\ref{app::approx_D_D}. When neglecting the space expansion term in appendix~\ref{app::space_term}, a freedom remained on $\chi$ from Eq.~\eqref{eq::mom_dec} as a spatial constant freedom. This constant is $H$. We took it to zero in appendix~\ref{app::space_term} in order to recover the 1+3-Newton equation as defined in Sec.~\ref{sec::Ortho_choice}. \\


As we included global expansion, the dictionary~\eqref{def::theta_dic_H}-\eqref{eq::dic_phys_R_ij_H} can be used to compare Newtonian and relativistic cosmological simulations. It is however still a bit limited as it requires irrotational fluids.

The modified 1+3-Newton equations of this section, and the related dictionary still do not feature backreaction in a compact space (the theorem of Buchert \& Ehlers still holds). Furthermore, the spatial sections being flat, we are not able to study structure formation in spherical or hyperbolic spaces, and the only oriented compact topology available is $\mathbb{T}^3$, up to a finite covering. In the next two subsections we will discuss possible GR based modifications of the 1+3-Newton equations which would allow these studies.

\subsubsection{Models for the study of the backreaction}

The Newton theory for fluid dynamics is a scalar-vector theory, i.e. the dynamical variables are a scalar and a vector. The scalar is the rotational free part of $\T\vN$ and the vector is the divergence free part of $\T\vN$. The scalar part is evolved with the Raychaudhuri equation~\eqref{eq::Ray_New4D_Orth} and the vector part with the vorticity equation~\eqref{eq::Wg_New4D_Orth}.

General relativity is a scalar-vector-tensor theory, i.e. there are dynamical variables, called tensorial variables, which cannot be written as function of a scalar or a vector. This is the case of the gravitational wave term in the decomposition~\eqref{eq::theta_decomp}.

What we mean by \textit{defining a GR based model from Newton's equations} is to keep the scalar-vector theory of Newton but with additional non-tensorial variables, terms and/or equations motivated by GR. Keeping a scalar-vector theory ensures a relative simplicity compared to tensor theories like GR. Such a model would enable the study of relativistic effects not present in Newton's theory using the simple tools of this theory.


In particular, we would like to focus on models implementing the backreaction which is a missing phenomenon of Newton's theory (for compact spaces). A possible model to study backreaction while allowing for non-linear structure formation would be to consider $\chi \not= 0$ in the decomposition~\eqref{eq::theta_decomp} along with $\D_\alpha \chi \not= 0$. As in Sec.~\ref{sec::Post_New_exp}, the expansion tensor features an additional term. But we consider here that $\chi$ is not a constant of space. The space expansion is thus local and global. For this model to be well defined one has to derive an evolution/constraint equation for the fundamental field $\chi$ from GR. This equation will feature $\T\vN$. Thus the Newtonian fluid dynamics will affect the space expansion. In this sense this model could be useful to probe the backreaction effect.

\subsubsection{For non-flat topologies}
\label{sec::New_non_flat}

In appendix~\ref{app::approx_D_D}, we assumed the spatial metric to be conformally flat to justify our Newton-GR dictionary. Relaxing this hypothesis and supposing $h_{ab}$ to be conformal to a constant curvature metric might be a way to define a Newtonian-like theory on a non-flat space.

Such a theory was heuristically defined in Ref.~\cite{2009_Roukema_et_al} to probe the topological acceleration in different spherical topologies. There were however different possibilities in this heuristic definition which were not relativistically motivated.

Adapting the limit leading to the 1+3-Newton formalism from GR (by changing the flat conformal hypothesis) could provide a non-flat Newtonian like theory coherent with general relativity. Along with the additional term $\chi$ in the expansion tensor, this theory if well defined, will be a tool to probe the effect of topology on the backreaction.

As an example, we give a possible model, but we do not try to justify it from GR. We consider, similarly to Ref.~\cite{2009_Roukema_et_al}, that the Newton equations~\eqref{eq::def_theta_omega_New}-\eqref{eq::g_def} are also valid if $\SNew$ is a constant curvature space\footnote{In Ref.~\cite{2009_Roukema_et_al}, the heuristic assumption regarding the curvature of $\SNew$ is made on the system~\eqref{eq::Cont_ENew}-\eqref{eq::Wg_ENew} which is not equivalent to doing it from the system~\eqref{eq::def_theta_omega_New}-\eqref{eq::g_def}.}, i.e. its Ricci tensor is $R_{ab} = \frac{R}{3} h_{ab}$, where $R$ is the scalar curvature. Then if we calculate the backreaction $\CQ_{\SNew}$, on the whole manifold $\SNew$, defined by Buchert \& Ehlers~\cite{1997_Buchert_et_al} as
\begin{align}
	\CQ_{\SNew} := {\left\langle \theta^2 - \Theta_{cd}\theta^{cd} +\omega_{cd}\omega^{cd}\right\rangle}_{\SNew} - \frac{2}{3}{\left\langle \theta \right\rangle}_{\SNew}^2,
\end{align}
it is not zero for a compact space (as for the flat case) anymore. Instead, we have the relation
\begin{align}
	\CQ_{\SNew} = \frac{R}{3} {\left\langle \vN_c\vN^c \right\rangle}_{\SNew} \label{eq::QD}
\end{align}
where the brackets ${\left\langle \cdot \right\rangle}_{\SNew}$ correspond to the spatial average over the compact space $\SNew$. This relation implies a dependence of the backreaction on the type of the global topology (spherical, flat or hyperbolic) via $\frac{R}{3}$, as well as on the Newtonian dynamics of the fluid via ${\left\langle \vN_c\vN^c \right\rangle}_{\SNew}$.

This model is however only heuristically defined, and thus we cannot be sure that the result~\eqref{eq::QD} is physically relevant. The 1+3-Newton formulation and the Newtonian limit we introduced might help justify, or disprove, this calculation.

\subsection{Comparison with the Newton-Cartan theory}

In the present section we explain the differences between our 4D formulation of Newton's equations and the Newton-Cartan theory.

The main difference is that we were able to define a non-degenerate metric on the spacetime manifold $\MNew$ of our formulation, implying this manifold to be (pseudo)-Riemannian, whereas this is not case in NC.

In addition to this point, the 1+3-Newton formulation does not feature an absolute time or foliation. We are free to change the foliation in which we are writing the equations. This leads to the 3+1-Newton equations (see Sec.~\ref{sec::3+1-Newton}). At most we can say that the formulation implies, like in GR, a preferred foliation: the one defined with respect to the fluid 4-velocity. The situation is different in NC where an absolute time is defined, linked to an absolute foliation.

Finally the 1+3-Newton system of equations~\eqref{eq::Cont_New4D_Orth}-\eqref{eq::def_theta_omega} is equivalent to the classical Newton system~\eqref{eq::def_theta_omega_New}-\eqref{eq::g_def}. The ensemble of solutions is then the same for both formulations. This is not the case in NC, where the theory is slightly more general than the classical formulation of Newton's theory (see Ref.~\cite{1997_Ehlers}).

\section{Conclusion}

The aim of this paper was to introduce a new formulation of Newton's equations on a 4-dimensional Lorentzian manifold.

To get to this formulation, we started from the classical Newton equations~\eqref{eq::Cont_ENew}-\eqref{eq::Wg_ENew} written in a Galilean frame on a 3-dimensional manifold $\SNew$. We generalised these equations by writing them for any time-parametrised coordinate system [Eqs.~\eqref{eq::Cont_NewP}-\eqref{eq::g_defP}]. We showed that the freedom on the choice of this coordinate system corresponds to a vector $\T \U$, defining what we called a class of coordinates $\classN{\T\U}$. This vector in general is not uniform, implying that its gradient is not zero. The symmetric part of the latter corresponds to the time variation of components of the metric [Eq.~\eqref{eq::part_h}], the antisymmetric part, if chosen to be a constant of space, corresponds to a global rotation of the coordinates $\classN{\T\U}$ with respect to a Galilean frame.

The freedom on $\T\U$ and the role it plays in the Newton equations~\eqref{eq::Cont_NewP}-\eqref{eq::g_defP} makes it very similar to the shift vector in the 3+1-formalism of general relativity. This allowed us to write the Newton equations as living in a 4-dimensional manifold $\MNew$. This was done using a push-forward on $\MNew$ of the classical Newton equations~\eqref{eq::Cont_ENew}-\eqref{eq::Wg_ENew} (see Sec.~\ref{sec::push_for}). The way the push-forward is done was inspired by the 3+1-Einstein equations. It was however not unique, which therefore implies that some freedom remains on the properties of $\MNew$. Regarding the signature freedom, we chose this manifold to be Lorentzian and argued that this was not in contradiction with Newton's theory (Sec.~\ref{sec::sig_New}). The remaining freedom (present as a lapse and a shift freedom) was chosen so that the foliation in which the equations are written corresponds to the rest frames of the Newtonian 4-velocity $\TuN$ we defined in Eq.~\eqref{eq::def_U_N_1}. This led to the \textit{1+3-Newton equations}~\eqref{eq::Cont_New4D_Orth}-\eqref{eq::def_theta_omega}. This set of equations is equivalent to the classical Newton equations, i.e. both can be derived from the other. This implies that the solutions described by the 1+3-Newton equations for the spatial fluid velocity $\T\vN$ are the same as the solutions of the classical Newton equations.

We then showed in Sec.~\ref{sec::3+1-Newton} that these equations, in the case of irrotational flows, can be recovered from general relativity in a limit $|\T\vN| \ll c$. This limit was performed with respect to a non-accelerating observer, the fluid itself. The Newtonian gravitational field $\T g$ then does not correspond to the 4-acceleration of a relativistic observer. Instead, it is defined as the acceleration, with respect to the fluid, of the spatial velocity $\T\vN$ [see Eq.~\eqref{eq::def_g_acc}]. The limit also showed what happens to the 3+1-Ricci equation of general relativity. This equation, not needed for a vector theory like Newton, is shown to be a relation for the second order rest frames curvature in the limit we introduced.

A first consequence of this limit is that the classical interpretation of the component $g_{00}$ of the spacetime metric as the gravitational potential still holds for coordinates adapted to a non-accelerating observer (Sec.~\ref{sec::New_limit}). This is however true only for stationary and irrotational fluids.

Another consequence of the limit is to define a dictionary (for irrotational flows) between the Newtonian fluid variables and general relativity (Sec.~\ref{sec::dico}). The spacetime manifold, denoted ${\cal M}^\mathrm{(dic)}$, given by this dictionary as function of the Newtonian variables is solution of the Einstein equations at leading order in $|\T\vN|/c$. In general ${\cal M}^\mathrm{(dic)} \not= \MNew$, implying that $\MNew$ is not solution of the Einstein equations at leading order. The difference between these two manifolds, ${\cal M}^\mathrm{(dic)}$ defined with the dictionary, and $\MNew$ defined with the 1+3-Newton equations, resides in the curvature of the fluid rest frames. It is exactly zero for $\MNew$, which is not the case for ${\cal M}^\mathrm{(dic)}$ where the curvature is of second order in $|\T\vN|/c$.

The dictionary was then tested for spherically symmetric vacuum solutions of Newton's and Einstein's theories of gravitation. For Newton this corresponded to a radially free-falling test fluid, and for general relativity to an observer associated with the generalised Gullstrand-Painlev\'e coordinates of the Schwarzschild spacetime. We showed that the dictionary allows us to recover the relativistic solution in the Newtonian limit. But in the specific case of a parabolic free-falling Newtonian fluid, the translation to general relativity is exact. This means that the Schwarzschild spacetime manifold is an exact solution, in terms of the manifold $\MNew$, of the 1+3-Newton equations. This supports our formulation. \\

The 1+3 formulation of Newton's equations might be seen as a new approach to evaluate the link between Newton's theory and general relativity. What is essentially new compared to other approaches (e.g.~\cite{2012_Green_et_al}) is that the comparison is done in the rest frames of the fluid, thus a non-accelerating observer. Furthermore we were able to construct a Lorentzian manifold on which the Newton equations are defined, contrasting with the Galilei manifold of the Newton-Cartan theory.

When developing this formulation we had in mind a future use for the study of the backreaction problem in cosmology and the effect of global topology. We think that the formulation might enable us to identify what is missing in Newton's theory for this study (due to the Buchert \& Ehlers theorem~\cite{1997_Buchert_et_al}, the backreaction is exactly zero for compact spaces in this theory). The final objective is then to use the 1+3-Newton equations and the scalar-vector-tensor decomposition of the expansion tensor in general relativity to define relatively simple models aimed at probing the backreaction and the effect of global topology. We give an example of what could be such a model in Sec.~\ref{sec::New_non_flat}.

Two things remain to be done before reaching this objective. The first one is to further test the dictionary for non-vacuum, non-stationary and non-isolated solutions. This can be done by comparing spherically symmetric solutions of Newton's equations with the Lemaitre-Tolman-Bondi class of solutions in general relativity. The second one is to upgrade the dictionary for vortical flows. We discussed this possibility in Sec.~\ref{sec::Dic_vort}.

\section*{Acknowledgements}

This work is part of a project that has received funding from the European Research Council (ERC) under the European Union’s Horizon 2020 research and innovation programme (Grant agreement ERC advanced Grant 740021–ARTHUS, PI: Thomas Buchert). I was supported by a ‘sp\'ecifique Normalien’ PhD Grant from the \'Ecole Normale Sup\'erieure de Lyon. I thank Pierre Mourier and Thomas Buchert for constant discussions and interest on this study and the manuscript. I would also like to thank \'Etienne Jaupart for essential debates on technical parts, L\'eo Brunswic for many valuable discussions and Boudewijn Roukema for comments on the manuscript.

\appendix

\section{Lie derivative in 3+1}
\label{app::Lie}

We consider a foliation $\{\Sigma_t\}_{t\in \mathbb{R}}$ in a manifold $\CM$. Let $\T A$ and $\T T$ be respectively a spatial vector and a spatial tensor. Then the Lie derivative of $\T T$ along $\T A$ is not necessarily spatial. It is spatial if $\T T$ has only contravariant indices. In general we have the relation
\begin{align}
	\Lie{\T A} {T^{\alpha_1 ...}}_{\beta_1 ...} = \	& \left({h^{\alpha_1}}_{\mu_1} ... \right) \left({h^{\nu_1}}_{\beta_1} ... \right) \Lie{\T A} {T^{\mu_1 ...}}_{\nu_1 ...} \\
										&- 2 \sum_{i} n_{\beta_i} {T^{\alpha_1 ...}}_{... \underset{\underset{i}{\uparrow}}{\nu} ...}  n_\mu \nabla^{(\mu}A^{\nu)}. \nonumber
\end{align}

Under the formalism of the 3+1-Einstein equations, given a class of adapted coordinates $\classn{\T \beta}$, this means that $\Lie{\T \beta} K_{\alpha\beta}$ is not a spatial tensor in general. This implies that ${^{\T \beta}\partial_t} K_{\alpha\beta}$ are not the spacetime components of a spatial tensor (contrary to what is stated in section 5.3.1 of Ref.~\cite{2012_GG}). However because the Lie derivative $\Lie{N\T n} = {^{\T \beta}\partial_t} - \Lie{\T \beta}$ applied on a spatial tensor is spatial, ${^{\T \beta}\partial_t} K_{\alpha\beta} - \Lie{\T \beta} K_{\alpha\beta}$ remains spatial. The pull-back ${^{\T \beta}\partial_t} K_{\alpha\beta} - \Lie{\T \beta} K_{\alpha\beta} \rightarrow {^{\T \beta}\partial_t} K_{ab} - \Lie{\T \beta} K_{ab}$ remains also true.

The normal part of the partial time derivative ${^{\T \beta}\partial_t} K_{\gamma\alpha}$ is
\begin{align}
	n^\gamma \ {^{\T \beta}\partial_t} K_{\gamma\alpha} = 4{K_\alpha}^\mu n^\nu \nabla_{(\mu}\beta_{\nu)}.
\end{align}

\section{Details on the approximations for the Newton-GR dictionary}
\label{app::approx_D_D}

We detail in this section arguments for the approximations made in Sec.~\ref{sec::hyp} regarding the decomposition of the expansion tensor and the covariant spatial derivative.

\subsection{Decomposition theorem?}
\label{app::decomp}

Equation~\eqref{eq::theta_decomp} is a decomposition only if each term in the RHS of this equation can be uniquely defined from  $\Theta_{ab}$. Straumann~\cite{2008_Straumann} showed that the decomposition~\eqref{eq::theta_decomp} for rank-2 tensors is always possible in compact\footnote{For non-compact spaces, fall-off conditions have to be used.} constant curvature spaces (constant scalar curvature and zero trace-less curvature). To our knowledge, no similar theorem exists for any curvature, which implies that the decomposition might be ill-defined in a general space. We however expect it to be reasonably valid for generally curved spaces in the context of cosmology. However the case of vortical flows remains problematic as no hypersurface orthogonal to the fluid 4-velocity can be defined.

\subsection{Approximation on the decomposition}

The only approximation which did not rely on $\beta_\vN~\ll~1$ regards Eq.~\eqref{eq::hyp} where we neglected the space expansion and the gravitational wave term. We will see in this section at which conditions it is consistent with a leading order approximation in $\beta_\vN$.

\subsubsection{Neglecting the space expansion}
\label{app::space_term}

Let us consider the momentum constraint~\eqref{eq::T_Codazzi_3D} with the decomposition~\eqref{eq::theta_decomp}. It becomes
\begin{align}
	\D_a\chi = \vN^cR_{ac}, \label{eq::mom_dec}
\end{align}
using $\D_{[a} \vN_{b]} = 0$, and the traceless and divergence free properties of the gravitational wave term.

We introduce the typical length scale $L_\chi$ of the space expansion. The RHS of Eq.~\eqref{eq::mom_dec} is of order $\beta_\vN^3/L_{\vN,l}^2$. So unless $L_{\vN,l}/L_\chi \gg 1$, which we assume is not the case for a cosmological setup, $\chi/c$ is at least of order $\beta_\vN^2/L_{\vN,l}$. Then neglecting the space expansion is coherent with a leading order approximation in $\beta_\vN$. \\

\remark{Actually this only shows that the space expansion term is a spatial constant at leading order. To recover the 1+3-Newton equation we take this constant to zero. However, letting it unspecified might be more interesting as it allows for expansion in a compact space (see Sec.~\ref{sec::Post_New_exp}).}

\subsubsection{Neglecting the gravitational wave term}

Let us consider that $\T\Xi = 0$ and $\chi \not= 0$, so that $\T\Theta = \chi\T h + \T\D\T\vN$. Then in the adapted coordinates $\class{-\T\vN}{\T u}$
\begin{align*}
	\derivt{-\T\vN}{\T u} h_{ab} = 2\chi h_{ab}.
\end{align*}
The solution to this differential equation can be written as
\begin{align}
	h_{ab} = \psi^2 \hh_{ab}, \label{eq::conform_h}
\end{align}
where $\hh_{ab}$ is called the \textit{background metric} with $\derivt{-\T\vN}{\T u} \hh_{ab} = 0$ and $\chi = \derivt{-\T\vN}{\T u}\ln\psi$. Note that this solution, while derived from a specific coordinate system, is however covariantly defined.

We introduce the covariant derivative $\TDh$ of $\Thh$. The conformal relation~\eqref{eq::conform_h} implies (see chapter 7 of Ref.~\cite{2012_GG} for details on this calculation\footnote{\label{ft::conform} In Ref.~\cite{2012_GG}, the conformal relation is $h_{ab} = \psi^4 \hh_{ab}$ instead of our $h_{ab} = \psi^2 \hh_{ab}$.})
\begin{align}
	R_{ab} =	\ & \Rh_{ab} - \Dh_a\Dh_b\ln\psi - \hat{h}_{ab} \Dh_c\Dh^c \ln\psi \label{eq::Ricci_confom} \\
			&+ \Dh_a\ln\psi \Dh_b \ln\psi - \hat{h}_{ab} \Dh_c\ln\psi \Dh^c \ln\psi, \nonumber
\end{align}
where $\TRh$ is the Ricci tensor relative to the metric $\Thh$, with $\derivt{-\T\vN}{\T u} \hat{R}_{ab} = 0$.

In general there does not exist a scalar $\psi$ and a time independent spatial metric $\hh_{ab}$ such that any Ricci tensor $R_{ab}$ can be written as in Eq.~\eqref{eq::Ricci_confom}. This equations is then a restriction for the form of $R_{ab}$ due to the initial assumption $\T\Xi = 0$. Consequently we expect the assumption $\Theta_{ab} \simeq \D_{a}\vN_{b}$ made in Sec.~\ref{sec::hyp} to \textit{not} be valid if, at leading order in $\beta_\vN$, the spatial Ricci tensor $\T R$ is not of the form of Eq.~\eqref{eq::Ricci_confom}. This can be seen as a test for the dictionary. \\

\subsection{Flat covariant derivative approximation}

We assume in this section that the spatial metric can be written as in Eq.~\eqref{eq::conform_h}, thus dropping the gravitational wave term. The Ricci tensor have the form~\eqref{eq::Ricci_confom}. This form suggests two typical curvature radius associated with $R_{ab}$: one for $\Rh_{ab}$ and one for the spatial variation of $\ln\psi$. We suppose the conformal metric $\hh_{ab}$ to be flat, so that $\TRh = 0$ and $\T R$ has only one typical radius as assumed in Sec.~\ref{sec::hyp}\footnote{Relaxing this hypothesis might lead to non-flat Newtonian-like theories as discussed in Sec.~\ref{sec::New_non_flat}.}. The relation $R_{ab} \sim 1/L_R^2$ implies that $\T\D\ln\psi \sim 1/L_R$.

The covariant derivative $\T\D$ can be written as function of $\psi$ and the covariant derivative $\TDh$ of $\hat{h}_{ab}$ (see chapter 7 of Ref.~\cite{2012_GG} for details on this calculation\footref{ft::conform}):
\begin{align}
	D_a\vN_b = \Dh_a\vN_b + h_{ab} \vN^c\Dh_c\ln\psi - 2 \, \vN_{(a}\Dh_{b)}\ln\psi. \label{eq::D_confom}
\end{align}
Using Eq.~\eqref{eq::L_v/L_R}, the second and third terms of the RHS of the decomposition~\eqref{eq::D_confom} are of the order $\beta_\vN^2$. So to first and leading order in $\beta_\vN$, we have $\T\D\T\vN = \TDh\T\vN$ which is what we assumed in the Newtonian limit of Sec.~\ref{sec::the_lim}.

\addcontentsline{toc}{section}{References}
\bibliography{paper_Newton}

\begin{thebibliography}{21}%
\makeatletter
\providecommand \@ifxundefined [1]{%
 \@ifx{#1\undefined}
}%
\providecommand \@ifnum [1]{%
 \ifnum #1\expandafter \@firstoftwo
 \else \expandafter \@secondoftwo
 \fi
}%
\providecommand \@ifx [1]{%
 \ifx #1\expandafter \@firstoftwo
 \else \expandafter \@secondoftwo
 \fi
}%
\providecommand \natexlab [1]{#1}%
\providecommand \enquote  [1]{``#1''}%
\providecommand \bibnamefont  [1]{#1}%
\providecommand \bibfnamefont [1]{#1}%
\providecommand \citenamefont [1]{#1}%
\providecommand \href@noop [0]{\@secondoftwo}%
\providecommand \href [0]{\begingroup \@sanitize@url \@href}%
\providecommand \@href[1]{\@@startlink{#1}\@@href}%
\providecommand \@@href[1]{\endgroup#1\@@endlink}%
\providecommand \@sanitize@url [0]{\catcode `\\12\catcode `\$12\catcode
  `\&12\catcode `\#12\catcode `\^12\catcode `\_12\catcode `\%12\relax}%
\providecommand \@@startlink[1]{}%
\providecommand \@@endlink[0]{}%
\providecommand \url  [0]{\begingroup\@sanitize@url \@url }%
\providecommand \@url [1]{\endgroup\@href {#1}{\urlprefix }}%
\providecommand \urlprefix  [0]{URL }%
\providecommand \Eprint [0]{\href }%
\providecommand \doibase [0]{http://dx.doi.org/}%
\providecommand \selectlanguage [0]{\@gobble}%
\providecommand \bibinfo  [0]{\@secondoftwo}%
\providecommand \bibfield  [0]{\@secondoftwo}%
\providecommand \translation [1]{[#1]}%
\providecommand \BibitemOpen [0]{}%
\providecommand \bibitemStop [0]{}%
\providecommand \bibitemNoStop [0]{.\EOS\space}%
\providecommand \EOS [0]{\spacefactor3000\relax}%
\providecommand \BibitemShut  [1]{\csname bibitem#1\endcsname}%
\let\auto@bib@innerbib\@empty
\bibitem [{\citenamefont {Ellis}(2009)}]{1971_Ellis}%
  \BibitemOpen
  \bibfield  {author} {\bibinfo {author} {\bibfnamefont {G.F.R.}\ \bibnamefont
  {Ellis}},\ }\bibfield  {title} {\enquote {\bibinfo {title} {{Relativistic
  cosmology}},}\ }\href {\doibase 10.1007/s10714-009-0760-7} {\bibfield
  {journal} {\bibinfo  {journal} {Gen.\ Rel.\ Grav.}\ }\textbf {\bibinfo
  {volume} {41}},\ \bibinfo {pages} {581--660} (\bibinfo {year}
  {2009})}\BibitemShut {NoStop}%
\bibitem [{\citenamefont {{Ellis}}(1967)}]{1967_Ellis}%
  \BibitemOpen
  \bibfield  {author} {\bibinfo {author} {\bibfnamefont {G.~F.~R.}\
  \bibnamefont {{Ellis}}},\ }\bibfield  {title} {\enquote {\bibinfo {title}
  {{Dynamics of Pressure-Free Matter in General Relativity}},}\ }\href
  {\doibase 10.1063/1.1705331} {\bibfield  {journal} {\bibinfo  {journal}
  {Journal of Mathematical Physics}\ }\textbf {\bibinfo {volume} {8}},\
  \bibinfo {pages} {1171--1194} (\bibinfo {year} {1967})}\BibitemShut {NoStop}%
\bibitem [{\citenamefont {{Buchert}}\ and\ \citenamefont
  {{Ehlers}}(1997)}]{1997_Buchert_et_al}%
  \BibitemOpen
  \bibfield  {author} {\bibinfo {author} {\bibfnamefont {T.}~\bibnamefont
  {{Buchert}}}\ and\ \bibinfo {author} {\bibfnamefont {J.}~\bibnamefont
  {{Ehlers}}},\ }\bibfield  {title} {\enquote {\bibinfo {title} {{Averaging
  inhomogeneous Newtonian cosmologies.}}}\ }\href@noop {} {\bibfield  {journal}
  {\bibinfo  {journal} {\aap}\ }\textbf {\bibinfo {volume} {320}},\ \bibinfo
  {pages} {1--7} (\bibinfo {year} {1997})},\ \Eprint
  {http://arxiv.org/abs/astro-ph/9510056} {arXiv:astro-ph/9510056 [astro-ph]}
  \BibitemShut {NoStop}%
\bibitem [{\citenamefont {{Sussman}}\ \emph {et~al.}(2016)\citenamefont
  {{Sussman}}, \citenamefont {{Delgado Gaspar}},\ and\ \citenamefont
  {{Hidalgo}}}]{2016_Sussman_et_al}%
  \BibitemOpen
  \bibfield  {author} {\bibinfo {author} {\bibfnamefont {Roberto~A.}\
  \bibnamefont {{Sussman}}}, \bibinfo {author} {\bibfnamefont {I.}~\bibnamefont
  {{Delgado Gaspar}}}, \ and\ \bibinfo {author} {\bibfnamefont {Juan~Carlos}\
  \bibnamefont {{Hidalgo}}},\ }\bibfield  {title} {\enquote {\bibinfo {title}
  {{Coarse-grained description of cosmic structure from Szekeres models}},}\
  }\href {\doibase 10.1088/1475-7516/2016/03/012} {\bibfield  {journal}
  {\bibinfo  {journal} {\jcap}\ }\textbf {\bibinfo {volume} {2016}},\ \bibinfo
  {eid} {012} (\bibinfo {year} {2016})},\ \Eprint
  {http://arxiv.org/abs/1507.02306} {arXiv:1507.02306 [gr-qc]} \BibitemShut
  {NoStop}%
\bibitem [{\citenamefont {{Giblin}}\ \emph {et~al.}(2016)\citenamefont
  {{Giblin}}, \citenamefont {{Mertens}},\ and\ \citenamefont
  {{Starkman}}}]{2016_Giblin_et_al_b}%
  \BibitemOpen
  \bibfield  {author} {\bibinfo {author} {\bibfnamefont {Jr.}\ \bibnamefont
  {{Giblin}}, \bibfnamefont {John~T.}}, \bibinfo {author} {\bibfnamefont
  {James~B.}\ \bibnamefont {{Mertens}}}, \ and\ \bibinfo {author}
  {\bibfnamefont {Glenn~D.}\ \bibnamefont {{Starkman}}},\ }\bibfield  {title}
  {\enquote {\bibinfo {title} {{Observable Deviations from Homogeneity in an
  Inhomogeneous Universe}},}\ }\href {\doibase 10.3847/1538-4357/833/2/247}
  {\bibfield  {journal} {\bibinfo  {journal} {\apj}\ }\textbf {\bibinfo
  {volume} {833}},\ \bibinfo {eid} {247} (\bibinfo {year} {2016})},\ \Eprint
  {http://arxiv.org/abs/1608.04403} {arXiv:1608.04403 [astro-ph.CO]}
  \BibitemShut {NoStop}%
\bibitem [{\citenamefont {{Macpherson}}\ \emph {et~al.}(2019)\citenamefont
  {{Macpherson}}, \citenamefont {{Price}},\ and\ \citenamefont
  {{Lasky}}}]{2019_Macpherson_et_al}%
  \BibitemOpen
  \bibfield  {author} {\bibinfo {author} {\bibfnamefont {Hayley~J.}\
  \bibnamefont {{Macpherson}}}, \bibinfo {author} {\bibfnamefont {Daniel~J.}\
  \bibnamefont {{Price}}}, \ and\ \bibinfo {author} {\bibfnamefont {Paul~D.}\
  \bibnamefont {{Lasky}}},\ }\bibfield  {title} {\enquote {\bibinfo {title}
  {{Einstein's Universe: Cosmological structure formation in numerical
  relativity}},}\ }\href {\doibase 10.1103/PhysRevD.99.063522} {\bibfield
  {journal} {\bibinfo  {journal} {\prd}\ }\textbf {\bibinfo {volume} {99}},\
  \bibinfo {eid} {063522} (\bibinfo {year} {2019})},\ \Eprint
  {http://arxiv.org/abs/1807.01711} {arXiv:1807.01711 [astro-ph.CO]}
  \BibitemShut {NoStop}%
\bibitem [{\citenamefont {{Brown}}(2009)}]{2009_Brown}%
  \BibitemOpen
  \bibfield  {author} {\bibinfo {author} {\bibfnamefont {J.~David}\
  \bibnamefont {{Brown}}},\ }\bibfield  {title} {\enquote {\bibinfo {title}
  {{Covariant formulations of Baumgarte, Shapiro, Shibata, and Nakamura and the
  standard gauge}},}\ }\href {\doibase 10.1103/PhysRevD.79.104029} {\bibfield
  {journal} {\bibinfo  {journal} {\prd}\ }\textbf {\bibinfo {volume} {79}},\
  \bibinfo {eid} {104029} (\bibinfo {year} {2009})},\ \Eprint
  {http://arxiv.org/abs/0902.3652} {arXiv:0902.3652 [gr-qc]} \BibitemShut
  {NoStop}%
\bibitem [{\citenamefont {{Buchert}}\ and\ \citenamefont
  {{Ostermann}}(2012)}]{RZA1}%
  \BibitemOpen
  \bibfield  {author} {\bibinfo {author} {\bibfnamefont {T.}~\bibnamefont
  {{Buchert}}}\ and\ \bibinfo {author} {\bibfnamefont {M.}~\bibnamefont
  {{Ostermann}}},\ }\bibfield  {title} {\enquote {\bibinfo {title} {{Lagrangian
  theory of structure formation in relativistic cosmology: Lagrangian framework
  and definition of a nonperturbative approximation}},}\ }\href {\doibase
  10.1103/PhysRevD.86.023520} {\bibfield  {journal} {\bibinfo  {journal}
  {Physical Review D}\ }\textbf {\bibinfo {volume} {86}},\ \bibinfo {eid}
  {023520} (\bibinfo {year} {2012})},\ \Eprint {http://arxiv.org/abs/1203.6263}
  {arXiv:1203.6263 [gr-qc]} \BibitemShut {NoStop}%
\bibitem [{\citenamefont {{Green}}\ and\ \citenamefont
  {{Wald}}(2012)}]{2012_Green_et_al}%
  \BibitemOpen
  \bibfield  {author} {\bibinfo {author} {\bibfnamefont {Stephen~R.}\
  \bibnamefont {{Green}}}\ and\ \bibinfo {author} {\bibfnamefont {Robert~M.}\
  \bibnamefont {{Wald}}},\ }\bibfield  {title} {\enquote {\bibinfo {title}
  {{Newtonian and relativistic cosmologies}},}\ }\href {\doibase
  10.1103/PhysRevD.85.063512} {\bibfield  {journal} {\bibinfo  {journal}
  {\prd}\ }\textbf {\bibinfo {volume} {85}},\ \bibinfo {eid} {063512} (\bibinfo
  {year} {2012})},\ \Eprint {http://arxiv.org/abs/1111.2997} {arXiv:1111.2997
  [gr-qc]} \BibitemShut {NoStop}%
\bibitem [{\citenamefont {{East}}\ \emph {et~al.}(2018)\citenamefont {{East}},
  \citenamefont {{Wojtak}},\ and\ \citenamefont {{Abel}}}]{2018_East_et_al}%
  \BibitemOpen
  \bibfield  {author} {\bibinfo {author} {\bibfnamefont {William~E.}\
  \bibnamefont {{East}}}, \bibinfo {author} {\bibfnamefont {Rados{\l}aw}\
  \bibnamefont {{Wojtak}}}, \ and\ \bibinfo {author} {\bibfnamefont {Tom}\
  \bibnamefont {{Abel}}},\ }\bibfield  {title} {\enquote {\bibinfo {title}
  {{Comparing fully general relativistic and Newtonian calculations of
  structure formation}},}\ }\href {\doibase 10.1103/PhysRevD.97.043509}
  {\bibfield  {journal} {\bibinfo  {journal} {\prd}\ }\textbf {\bibinfo
  {volume} {97}},\ \bibinfo {eid} {043509} (\bibinfo {year} {2018})},\ \Eprint
  {http://arxiv.org/abs/1711.06681} {arXiv:1711.06681 [astro-ph.CO]}
  \BibitemShut {NoStop}%
\bibitem [{\citenamefont {{K{\"u}nzle}}(1976)}]{1976_Kunzle}%
  \BibitemOpen
  \bibfield  {author} {\bibinfo {author} {\bibfnamefont {H.~P.}\ \bibnamefont
  {{K{\"u}nzle}}},\ }\bibfield  {title} {\enquote {\bibinfo {title} {{Covariant
  Newtonian limit of Lorentz space-times}},}\ }\href {\doibase
  10.1007/BF00766139} {\bibfield  {journal} {\bibinfo  {journal} {General
  Relativity and Gravitation}\ }\textbf {\bibinfo {volume} {7}},\ \bibinfo
  {pages} {445--457} (\bibinfo {year} {1976})}\BibitemShut {NoStop}%
\bibitem [{\citenamefont {{Ehlers}}(2019)}]{2019_Ehlers}%
  \BibitemOpen
  \bibfield  {author} {\bibinfo {author} {\bibfnamefont {J.}~\bibnamefont
  {{Ehlers}}},\ }\bibfield  {title} {\enquote {\bibinfo {title} {{Republication
  of: On the Newtonian limit of Einstein’s theory of gravitation}},}\ }\href
  {\doibase 10.1007/s10714-019-2624-0} {\bibfield  {journal} {\bibinfo
  {journal} {General Relativity and Gravitation}\ }\textbf {\bibinfo {volume}
  {51}} (\bibinfo {year} {2019}),\ 10.1007/s10714-019-2624-0}\BibitemShut
  {NoStop}%
\bibitem [{\citenamefont {{Gourgoulhon}}(2012)}]{2012_GG}%
  \BibitemOpen
  \bibfield  {author} {\bibinfo {author} {\bibfnamefont {E.}~\bibnamefont
  {{Gourgoulhon}}},\ }\href {\doibase 10.1007/978-3-642-24525-1} {\emph
  {\bibinfo {title} {{3+1 Formalism in General Relativity: Bases of Numerical
  Relativity}}}},\ \bibinfo {series} {Lecture Notes in Physics}, Vol.\ \bibinfo
  {volume} {846}\ (\bibinfo  {publisher} {Springer-Verlag Berlin Heidelberg},\
  \bibinfo {year} {2012})\BibitemShut {NoStop}%
\bibitem [{\citenamefont {{Roy}}(2014)}]{2014_Roy}%
  \BibitemOpen
  \bibfield  {author} {\bibinfo {author} {\bibfnamefont {Xavier}\ \bibnamefont
  {{Roy}}},\ }\bibfield  {title} {\enquote {\bibinfo {title} {{On the 1+3
  Formalism in General Relativity}},}\ }\href@noop {} {\bibfield  {journal}
  {\bibinfo  {journal} {arXiv e-prints}\ ,\ \bibinfo {eid} {arXiv:1405.6319}}
  (\bibinfo {year} {2014})},\ \Eprint {http://arxiv.org/abs/1405.6319}
  {arXiv:1405.6319 [gr-qc]} \BibitemShut {NoStop}%
\bibitem [{\citenamefont {{Maartens}}\ and\ \citenamefont
  {{Bassett}}(1998)}]{1998_Maartens_et_al}%
  \BibitemOpen
  \bibfield  {author} {\bibinfo {author} {\bibfnamefont {Roy}\ \bibnamefont
  {{Maartens}}}\ and\ \bibinfo {author} {\bibfnamefont {Bruce~A.}\ \bibnamefont
  {{Bassett}}},\ }\bibfield  {title} {\enquote {\bibinfo {title}
  {{Gravito-electromagnetism}},}\ }\href {\doibase 10.1088/0264-9381/15/3/018}
  {\bibfield  {journal} {\bibinfo  {journal} {Classical and Quantum Gravity}\
  }\textbf {\bibinfo {volume} {15}},\ \bibinfo {pages} {705--717} (\bibinfo
  {year} {1998})},\ \Eprint {http://arxiv.org/abs/gr-qc/9704059}
  {arXiv:gr-qc/9704059 [gr-qc]} \BibitemShut {NoStop}%
\bibitem [{\citenamefont {{Vincent}}\ \emph {et~al.}(2012)\citenamefont
  {{Vincent}}, \citenamefont {{Gourgoulhon}},\ and\ \citenamefont
  {{Novak}}}]{2012_Vincent_et_al}%
  \BibitemOpen
  \bibfield  {author} {\bibinfo {author} {\bibfnamefont {F.~H.}\ \bibnamefont
  {{Vincent}}}, \bibinfo {author} {\bibfnamefont {E.}~\bibnamefont
  {{Gourgoulhon}}}, \ and\ \bibinfo {author} {\bibfnamefont {J.}~\bibnamefont
  {{Novak}}},\ }\bibfield  {title} {\enquote {\bibinfo {title} {{3+1 geodesic
  equation and images in numerical spacetimes}},}\ }\href {\doibase
  10.1088/0264-9381/29/24/245005} {\bibfield  {journal} {\bibinfo  {journal}
  {Classical and Quantum Gravity}\ }\textbf {\bibinfo {volume} {29}},\ \bibinfo
  {eid} {245005} (\bibinfo {year} {2012})},\ \Eprint
  {http://arxiv.org/abs/1208.3927} {arXiv:1208.3927 [gr-qc]} \BibitemShut
  {NoStop}%
\bibitem [{\citenamefont {{MacLaurin}}(2019)}]{2019_MacLaurin}%
  \BibitemOpen
  \bibfield  {author} {\bibinfo {author} {\bibfnamefont {Colin}\ \bibnamefont
  {{MacLaurin}}},\ }\bibfield  {title} {\enquote {\bibinfo {title}
  {{Schwarzschild spacetime under generalised Gullstrand-Painlev{\'e}
  slicing}},}\ }\href@noop {} {\bibfield  {journal} {\bibinfo  {journal} {arXiv
  e-prints}\ ,\ \bibinfo {eid} {arXiv:1911.05988}} (\bibinfo {year} {2019})},\
  \Eprint {http://arxiv.org/abs/1911.05988} {arXiv:1911.05988 [gr-qc]}
  \BibitemShut {NoStop}%
\bibitem [{\citenamefont {{Buchert}}\ \emph {et~al.}(2020)\citenamefont
  {{Buchert}}, \citenamefont {{Mourier}},\ and\ \citenamefont
  {{Roy}}}]{2020_BMR}%
  \BibitemOpen
  \bibfield  {author} {\bibinfo {author} {\bibfnamefont {Thomas}\ \bibnamefont
  {{Buchert}}}, \bibinfo {author} {\bibfnamefont {Pierre}\ \bibnamefont
  {{Mourier}}}, \ and\ \bibinfo {author} {\bibfnamefont {Xavier}\ \bibnamefont
  {{Roy}}},\ }\bibfield  {title} {\enquote {\bibinfo {title} {{On average
  properties of inhomogeneous fluids in general relativity III: general fluid
  cosmologies}},}\ }\href {\doibase 10.1007/s10714-020-02670-6} {\bibfield
  {journal} {\bibinfo  {journal} {General Relativity and Gravitation}\ }\textbf
  {\bibinfo {volume} {52}},\ \bibinfo {eid} {27} (\bibinfo {year} {2020})},\
  \Eprint {http://arxiv.org/abs/1912.04213} {arXiv:1912.04213 [gr-qc]}
  \BibitemShut {NoStop}%
\bibitem [{\citenamefont {{Roukema}}\ and\ \citenamefont
  {{R{\'o}{\.z}a{\'n}ski}}(2009)}]{2009_Roukema_et_al}%
  \BibitemOpen
  \bibfield  {author} {\bibinfo {author} {\bibfnamefont {B.~F.}\ \bibnamefont
  {{Roukema}}}\ and\ \bibinfo {author} {\bibfnamefont {P.~T.}\ \bibnamefont
  {{R{\'o}{\.z}a{\'n}ski}}},\ }\bibfield  {title} {\enquote {\bibinfo {title}
  {{The residual gravity acceleration effect in the Poincar{\'e} dodecahedral
  space}},}\ }\href {\doibase 10.1051/0004-6361/200911881} {\bibfield
  {journal} {\bibinfo  {journal} {\aap}\ }\textbf {\bibinfo {volume} {502}},\
  \bibinfo {pages} {27--35} (\bibinfo {year} {2009})},\ \Eprint
  {http://arxiv.org/abs/0902.3402} {arXiv:0902.3402 [astro-ph.CO]} \BibitemShut
  {NoStop}%
\bibitem [{\citenamefont {{Ehlers}}(1997)}]{1997_Ehlers}%
  \BibitemOpen
  \bibfield  {author} {\bibinfo {author} {\bibfnamefont {J{\"u}rgen}\
  \bibnamefont {{Ehlers}}},\ }\bibfield  {title} {\enquote {\bibinfo {title}
  {{Examples of Newtonian limits of relativistic spacetimes}},}\ }\href
  {\doibase 10.1088/0264-9381/14/1A/010} {\bibfield  {journal} {\bibinfo
  {journal} {Classical and Quantum Gravity}\ }\textbf {\bibinfo {volume}
  {14}},\ \bibinfo {pages} {A119--A126} (\bibinfo {year} {1997})}\BibitemShut
  {NoStop}%
\bibitem [{\citenamefont {{Straumann}}(2008)}]{2008_Straumann}%
  \BibitemOpen
  \bibfield  {author} {\bibinfo {author} {\bibfnamefont {N.}~\bibnamefont
  {{Straumann}}},\ }\bibfield  {title} {\enquote {\bibinfo {title} {{Proof of a
  decomposition theorem for symmetric tensors on spaces with constant
  curvature}},}\ }\href {\doibase 10.1002/andp.200810312} {\bibfield  {journal}
  {\bibinfo  {journal} {Annalen der Physik}\ }\textbf {\bibinfo {volume}
  {520}},\ \bibinfo {pages} {609--611} (\bibinfo {year} {2008})},\ \Eprint
  {http://arxiv.org/abs/0805.4500} {arXiv:0805.4500 [gr-qc]} \BibitemShut
  {NoStop}%
\end{thebibliography}%

\end{document}